2024

# Exploring Fraction Comprehension and Interest in Elementary Education Through AI-Powered Personalized Learning

Kenneth Holman
*University of Central Florida*

# EXPLORING FRACTION COMPREHENSION AND INTEREST IN ELEMENTARY EDUCATION THROUGH AI-POWERED PERSONALIZED LEARNING

by

KENNETH ANDREW HOLMAN II
B.S. University of Central Florida, 2005

M.Ed. Bethune Cookman University, 2020

A dissertation submitted in partial fulfillment of the requirements
for the degree of Doctor of Philosophy
in the Department of Learning Sciences and Educational Research
in the College of Community Innovation and Education
at the University of Central Florida
Orlando, Florida

Fall Term
2024

Major Professor: Matthew T. Marino

# ABSTRACT

Fraction proficiency is crucial for foundational mathematics and broader STEM fields. This dissertation investigates the integration of Artificial Intelligence (AI) in mathematics classrooms, with a primary focus on students with mathematical learning challenges. AI technologies offer unique advantages for personalizing instruction and providing tailored feedback. The empirical study in the dissertation addresses a gap in the research related to AI's application during primary school fraction instruction.

The first manuscript presents a comprehensive literature review of AI in mathematics education, synthesizing research from 2020 - 2024. The second manuscript describes a quasi-experimental study evaluating Mathbot, an AI-based personalized learning platform. Repeated Measures ANOVA revealed modest improvements in fraction comprehension for students using Mathbot compared to traditional instruction, though changes in situational interest were not statistically significant. The third manuscript explores the application of AI-powered personalized learning in special education, providing strategies for pre-service teachers.

This dissertation illustrates AI's potential in mathematics education, while highlighting the need for further research to evaluate its effectiveness. It offers insights related to leveraging AI-driven learning to enhance outcomes for students, particularly those with learning challenges. Findings included herein will contribute to the evolving landscape of AI in special education.

Karena, I dedicate this dissertation to you. It would not have been possible without your continued support and love through this process. You are my rock and keep me safe when I am falling. You were always there; if it were up to me, you would walk across the stage because this process was as difficult for you as myself. I was not always "present" when I needed to be during this process, but I know you were one of the only reasons I could accomplish this goal. I love you forever and always.

Caleb, Maya Gwen, and Ro, so you all know, much of this was done because of you. I always want to show you that no matter what, you can accomplish any goal you set your mind to. I hope I have set a positive example for you showing that you can accomplish whatever you set your mind to no matter your lot in life. I feel like I did a decent job supporting you three through this process, making sure I was at every event, no matter how small, whether it be an awards ceremony, sporting event, or even just going to starbies. I love all three of you more than you could possibly imagine.

Ma and Pop, your support is very much appreciated. I was never concerned about my family doing without because of you. I always knew that no matter what, I had people I could depend on; you are and will always be my biggest cheerleaders. I love you both.

## ACKNOWLEDGEMENTS

Christine, you have been there through this whole process, and for that, I am thankful. If I ever needed anything, you would guide, support, and mostly tell me it's all going to be OK while being honest and not letting me off the hook. Dr. Taub, you are one of the kindest, most gentle human beings I have ever met, plus you are a rock star in an area I am not. I appreciate you taking time, even when you didn't have to help through this process. Dr. Hunt, you have shown me it's ok to be me and do what makes me happy; don't settle and mentor and guide me through this process; you have seen me grow, and I appreciate you. Dr. Vasquez, although you intimidate me, which I think you enjoy, you have shown me always to try to think outside the box and innovate. It is because of you I don't settle for the status quo and always attempt to push the science and move the needle. I will forever be grateful for your mentorship and guidance. Dr. Marino, you have let me research and learn in a way which I was comfortable. You are hands-off and expect a lot of me, but I have grown more than you could imagine. My beliefs, goals, and ambitions have changed during this process because of you. You knew when I needed a swift kick and would not hesitate to tell me, and I am extremely grateful for your guidance and mentorship.

**TABLE OF CONTENTS**

# LIST OF FIGURES

## LIST OF TABLES

## LIST OF ACRONYMS

| | |
|---|---|
| AI | Artificial Intelligence |
| AI-PPL | Artificial Intelligence-Powered Personalized Learning |
| BAU | Business-As-Usual |
| DGBLE | Digital Game-Based Learning Environment |
| GMM | Gaussian Mixture Model |
| FCA | Fraction Comparison Assessment |
| ITS | Intelligent Tutoring System |
| MLC | Mathematical Learning Challenges |
| RMANOVA | Repeated Measures Analysis of Variance |
| SIS | Situational Interest Survey |
| STEM | Science, Technology, Engineering, and Mathematics |
| UDL | Universal Design for Learning |

# CHAPTER ONE: INTRODUCTION

Fractions are foundational within the broader mathematics curriculum, facilitating the transition to more advanced concepts such as algebra, statistics, and calculus (Wortha et al., 2023). Mastery of fractions is essential for both academic success and everyday tasks like cooking, managing finances, understanding interest rates, and interpreting parts of a whole. Fraction knowledge is essential for functional numeracy, allowing individuals to solve real-world problems more accurately and efficiently (Handel, 2016). Recent studies underscore that mastering fractions strongly predicts success in higher mathematics and, more broadly, STEM fields (Perry, 2023; Theobald et al., 2020).

Given their importance in both academic and practical contexts, strong fraction comprehension should be emphasized early in education. Mastering fractions enhances problem-solving abilities, including logical reasoning. It establishes a foundation for future vocational and academic STEM success (Spitzer & Moeller, 2021). However, students with mathematical learning challenges are struggling more now than in the past.

The COVID-19 pandemic intensified educational disparities, accelerating learning losses in mathematics across all grade levels. According to the National Center for Educational Statistics (2022), only 36% of 4th-grade students demonstrated proficiency in mathematics, marking a significant decline in performance. This trend is especially pronounced in fraction comprehension, which saw its first notable decline since 2015 among students with and without disabilities (Fuchs et al., 2023; Kuhfeld et al., 2022). These troubling statistics underscore the urgent need for targeted interventions to halt the decline in mathematical achievement, particularly in fractions.

## The Need for AI Research in Mathematics Education

Artificial intelligence (AI) technologies offer promising solutions for these challenges (Marino et al., 2023). AI-powered tools, such as intelligent tutoring systems and adaptive learning platforms, can identify students' misconceptions and provide tailored feedback in real-time (Bush, 2021). These tools dynamically adjust instructional content based on individual learning patterns, allowing for personalized and interactive learning experiences (Capinding, 2023). Mathbot, an AI-based personalized learning platform, exemplifies such technology by integrating students' interests into fraction problems, creating an engaging and customized approach to mathematics education. Research suggests that AI-driven systems like Mathbot have the potential to bridge the gap in fraction comprehension, particularly for students facing mathematical learning challenges (Satsangi & Raines, 2022).

Despite the promising capabilities of AI technologies, research on their application in fraction education remains limited (Hwang & Tu, 2021). Most existing studies focus on AI's role in broad mathematics education, without focusing on fraction comprehension (Mohamed et al., 2022). Researchers have identified ethical considerations, such as data privacy, algorithmic bias, and the need for culturally inclusive AI tools, as significant areas requiring further exploration (Vasquez et al., 2024). This study addresses these gaps by exploring how AI-powered tools enhance fraction comprehension in elementary students. Specifically, the study focuses on Mathbot, an AI-based personalized learning platform that dynamically adapts to individual learning patterns and integrates personal interests into fraction problems, offering a novel approach to fraction instruction.

## Theoretical Underpinnings of AI Interventions

The implementation of AI in mathematics education is grounded in theories of personalized learning, which suggest that adapting educational content to meet individual learners' needs can enhance comprehension and interest. By personalizing the learning experience, AI tools like Mathbot can support students with mathematical learning challenges, addressing both academic and motivational needs, as highlighted in various studies within the literature review (Annuš & Kmeť, 2024; Capinding, 2023). Additionally, AI aligns with direct instruction by providing structured guidance and immediate feedback. AI-powered tools can offer explicit teaching methods, such as step-by-step explanations of mathematical concepts, while also adapting to individual students' needs in real time, which mimics traditional teacher-led instruction (Martin & Hunt, 2022; Akavova et al., 2023). This approach helps reinforce foundational skills, ensuring students can master essential concepts before advancing to more complex material. Furthermore, AI resonates with constructivist theories by promoting active engagement and personalized problem-solving experiences. Through adaptive learning platforms, AI allows students to build new knowledge based on their prior understanding, supporting deeper comprehension, and encouraging students to take an active role in their learning (Daghestani et al., 2020; Pappas & Drigas, 2016). By integrating direct instruction with constructivist elements, AI enhances learning by personalizing both the pace and method of instruction to suit individual students.

The Mathbot intervention draws from both Papert's Constructionism and Skinner's Reinforcement Learning Theory, forming a foundation for adaptive AI-driven learning environments. Papert's Constructionism emphasizes hands-on, active learning through interaction and reflection. Mathbot operationalizes these principles by allowing students to engage directly with fraction problems, providing immediate feedback to help them reflect on and refine their understanding. This aligns with Papert's view that learners construct knowledge most effectively through active engagement and problem-solving rather than passively receiving information (Papert, 1985). Group discussions after Mathbot sessions further reinforce these principles, encouraging reflection and peer-to-peer learning, which deepens understanding and aids in the construction of knowledge.

In contrast, Skinner's Reinforcement Learning Theory focuses on shaping behavior through immediate feedback and reinforcement. In the Mathbot system, students receive positive reinforcement for correct answers, which encourages the repetition of successful problem-solving strategies. Incorrect answers prompt additional scaffolding and practice, guiding students toward mastery (Skinner, 1953). This feedback loop reflects Skinner's belief that timely reinforcement strengthens desired learning behaviors, promoting effective and continuous learning. Mathbot's ability to immediately address misconceptions and offer additional practice makes it a valuable tool in shaping students' learning behaviors over time.

Constructionism and reinforcement learning complement each other, balancing exploratory learning with behavioral reinforcement. While Papert's model encourages exploration and interaction, Skinner's framework ensures that learning is directed and reinforced, leading to measurable improvements in student outcomes. Mathbot's design capitalizes on this association, providing students with both the freedom to explore mathematical concepts and the structured feedback necessary to guide their learning process.

## Purpose and Structure of the Dissertation

The objectives of this dissertation were to examine current AI research in mathematics education (Chapter 2), develop and test the Mathbot AI intervention (Chapter 3), and develop a set of guiding principles for preservice teachers who are interested in using AI in their math classrooms (Chapter 4).

# Chapter 2: Artificial Intelligence in Mathematics Education: A Systematic Literature Review

The literature review systematically explored the integration and effectiveness of Artificial Intelligence (AI) in mathematics education, focusing on AI-driven tools and methodologies used in classroom interventions. What makes this review unique is its comprehensive scope, examining 14 studies involving 8,305 participants across diverse educational settings—primary, secondary, and post-secondary. This extensive dataset allows the review to provide a more nuanced understanding of AI's impact on mathematics education. The inclusion of various age groups and academic levels offers insights into the scalability and adaptability of AI tools like Intelligent Tutoring Systems (ITS) and adaptive learning platforms for different learners. These technologies facilitate personalized learning by dynamically adapting to individual student needs, improving engagement, and fostering academic success.

Additionally, the review identifies the specific benefits of AI in handling complex mathematical concepts, such as fractions and algebra, where real-time feedback from AI tools has shown to be particularly effective (Bush, 2021; Capinding, 2023). Another unique aspect of this review is its emphasis on the ethical considerations associated with AI in educational settings, including the critical issues of data privacy, algorithmic bias, and transparency in decision-making processes. These concerns are crucial for equitable AI deployment in schools (Baker & Hawn, 2021; Sperling et al., 2022). The attention to these ethical concerns makes this review forward-looking, as it underscores the need for ongoing monitoring of AI's role in education to mitigate potential risks while maximizing its benefits.

A significant contribution of this review lies in the clear trend it identifies regarding the link between the duration of AI-based interventions and their effect sizes. The review highlights how longer interventions, spanning weeks or months, yield larger effect sizes, such as those seen in the "Learn with ME" program (Annuš & Kmeť, 2024) and Photomath (Capinding, 2023). In contrast, shorter interventions, like those examined by Shih et al. (2023) and Cohen et al. (2021), produce smaller effect sizes, emphasizing the importance of sustained interaction with AI tools to realize their full potential (de Boer et al., 2014; Foster, 2023). This connection between duration and effectiveness distinguishes this review, as it calls for longer-term strategies to ensure meaningful and lasting improvements in mathematical comprehension and skills. This review not only synthesizes the effectiveness of AI tools in mathematics education but also highlights critical factors such as ethical deployment and duration of use, offering a roadmap for future research to refine and optimize AI interventions in education.

## Chapter 3: Assessing Mathbot's Impact on Fraction Comprehension

This chapter examined the effectiveness of Mathbot, an AI-powered personalized learning tool designed to improve fraction comprehension and situational interest among elementary students with mathematical learning challenges (MLC). The intervention was conducted over five consecutive school days, with students participating in 40-minute sessions during their regular math periods. The sample consisted of 22 fourth and fifth graders from a suburban school in the southeastern United States. The participants were divided into two groups: the Mathbot group, which utilized the AI tool for personalized learning, and a business-as-usual (BAU) control group, which followed traditional teacher-led instruction. Pre-test and post-test assessments were administered to measure changes in fraction comprehension using the Fraction Comparison Assessment (FCA), and situational interest was measured using a six-item Situational Interest Survey (SIS). Teachers in both groups followed the same lesson plan focused on fraction comparison to ensure consistency, while the Mathbot group, powered by AI, received adaptive feedback personalized to their learning needs. The study employed a quasi-experimental design with repeated measures, and data were analyzed using a repeated measures ANOVA to compare the outcomes between the two groups (Mathbot and BAU) over time.

The results indicated that while both groups demonstrated gains in fraction comprehension, the BAU group slightly outperformed Mathbot users in post-test assessments. However, no statistically significant improvements were observed in situational interest for either group. These findings highlight the potential of AI tools like Mathbot in supporting fraction comprehension, but also indicate that more development is needed to sustain student interest in mathematics. The study highlights the potential and limitations of AI-powered educational tools and calls for the continued refinement of AI interventions to fully realize their potential in fostering academic and motivational outcomes in personalized learning environments.

## Chapter 4: Navigating AI-Powered Personalized Learning in Special Education: A Guide for Preservice Teacher Faculty

This chapter explored the integration of AI-Powered Personalized Learning (AI-PPL) in special education, particularly for preservice teachers. AI-PPL tools offer tailored learning experiences and adaptive feedback, enhancing engagement for students with disabilities, especially in STEM-related subjects. This chapter provides educators with practical strategies for incorporating AI into special education environments. AI systems like Mathbot offer personalized content based on student's interests and abilities, making learning more accessible and motivating. These tools track progress and provide instant feedback, allowing educators to intervene when necessary. The chapter emphasizes the importance of professional development, highlighting teachers' need to gain technical and pedagogical expertise to implement AI tools effectively. Ethical considerations ensuring accessibility and inclusivity are crucial when integrating AI in educational settings. Overall, AI-PPL offers a transformative approach to personalized learning, making it possible to meet the diverse needs of students with disabilities while fostering an inclusive and supportive educational environment.

This chapter builds on insights from Chapters 2 and 3, which examined AI's use in mathematics education and Mathbot's impact on fraction comprehension. The exploration of AI-Powered Personalized Learning (AI-PPL) in special education further expands this conversation, emphasizing its critical role in addressing diverse student needs. The potential impact of AI in education, particularly for students with disabilities, is profound. By offering adaptive learning environments tailored to individual capabilities, AI can transform traditional special education, shifting from a one-size-fits-all model to one that is flexible and personalized. This has implications not only for academic achievement but also for social inclusion and motivation, as AI tools can foster more engaging and supportive educational experiences. Additionally, the effective integration of AI in preservice teacher training, as outlined in this chapter, holds the potential to reshape how future educators understand and implement personalized learning strategies. By equipping teachers with the technical and pedagogical skills necessary to utilize AI, we can ensure that AI tools are leveraged to their full potential, creating more equitable and accessible educational environments for all learners.

# CHAPTER TWO: SYSTEMATICS REVIEW: ARTIFICIAL INTELLIGENCE IN MATHEMATICS EDUCATION

This chapter titled “Artificial Intelligence in Mathematics Education: A Systematic Literature Review” will be submitted for publication in the referred professional journal titled, *International Electronic Journal of Mathematics Education*.

Abstract

This systematic literature review examines the integration of Artificial Intelligence (AI) in mathematics education, focusing on tools such as Intelligent Tutoring Systems (ITS), adaptive learning platforms, and AI-powered educational apps in education settings ranging from elementary to post-secondary schools. The review assesses the impact of these technologies on student learning in 14 studies from 2020 to 2024. AI-based tools enhanced comprehension, engagement, problem-solving, and critical thinking by personalizing learning experiences. However, significant challenges remain, including data privacy concerns, algorithmic bias, and the need for professional development. While AI was effective for improving basic problem-solving, it required teacher mediation to handle more complex tasks. Outcomes from this review provide evidence of the need for further research examining AI integration to support diverse learning needs.
*Keywords*: artificial intelligence, mathematics education, intelligent tutoring systems (ITS), adaptive learning platforms, AI-powered educational apps, personalized learning, student engagement, educational outcomes, data privacy.

## Introduction

Mathematical principles are foundational to modern society, supporting everything from practical tasks to data analysis (Sharma, 2021). A solid foundation in mathematics is essential for educational and career success (Haigh, 2019). Yet, only 36% of fourth and 26% of eighth graders are proficient in mathematics, with proficiency rates dropping for students with disabilities, at just 16% and 7% respectively (National Center for Education Statistics, 2022). The COVID-19 pandemic exacerbated these challenges, leading to the first decline in national math scores since 2015 (Fuchs et al., 2023; Kuhfeld et al., 2022).

Innovative teaching strategies, such as integrating video game-based learning, have shown promise for improving learning outcomes when students learn mathematical concepts such as fractions (Hunt et al., 2023). Similarly, artificial intelligence (AI) technologies have been shown to enhance comprehension and engagement during engineering education (Jian, 2023). AI's ability to provide real-time feedback, adjust learning paths, and support in-class and remote learning make it a versatile tool (Rakya, 2023). AI has the potential to address learning deficits exposed during the pandemic (Saravanan et al., 2022). A meta-analysis by Hwang (2022) noted AI could bridge achievement gaps in mathematics education. However, additional research is necessary (Basham et al., 2024).

## History of AI in Education

AI began with Alan Turing's pioneering work in the mid-20th century. In his seminal paper, *Computing Machinery and Intelligence* (1950), Turing introduced the Turing Test, a method to evaluate whether a machine could exhibit intelligent behavior indistinguishable from a human (Turing, 1950). This concept challenged conventional views of machines and laid the foundation for modern AI. Turing's development of the Turing machine, an abstract model capable of simulating any algorithmic process, was a landmark in computer science, setting the stage for future AI research (Bowen, 2016).

Building on Turing's work, machine learning algorithms emerged in the late 1950s, becoming central to modern AI (Samuel, 1959). During this time, researchers explored neural networks inspired by biological neurons to mimic human decision-making processes (Rosenblatt, 1958). Early AI systems were designed to learn from data, which led to the development of adaptive algorithms in computing (Hall, 2007). These advancements were pivotal for educational applications, particularly in developing adaptive learning systems tailored to individual student needs (Ullman, 2019).

By the 1980s, AI's impact on education became more prominent with the development of systems that delivered personalized learning experiences. Intelligent Tutoring Systems (ITS) emerged, employing AI to adapt to students' learning speeds, styles, and needs (Mozer et al., 2019). These systems provided students with personalized instructional feedback in real-time, significantly enhancing educational outcomes (Kochmar et al., 2020). ITS marked a major shift in educational technologies, laying the foundation for modern personalized learning environments. These systems adapted content delivery to each student's needs, laying the groundwork for today's more sophisticated personalization and adaptive technologies (Brynjolfsson, 2022).

In the early 21st century, AI's role in education broadened to encompass sophisticated educational assessments (González-Calatayud et al., 2021). AI systems began analyzing student responses and behaviors, providing educators deeper insights into how students learn and solve problems (Owan et al., 2023). Innovations such as data mining and natural language processing enabled AI to assess the correctness of student answers and their thought processes (Shaik et al., 2023; Liang et al., 2021). This transformed traditional educational assessments, facilitating automated short-answer grading, knowledge tracing, and class quality assurance (Liu et al., 2020).

In 2023, the U.S. Department of Education released a report outlining AI's role in enhancing education, focusing on improving learning outcomes through personalization and addressing system-level risks. The report emphasized the need for ethical, equitable, and transparent AI models to protect student privacy and mitigate biases in educational settings. Subsequently, 19 states established AI educational policies, providing guidance focused on K-12 education and personalized learning for all students, including those with disabilities (CIDDL, 2024). However, as Marino et al. (2023) noted, additional research is necessary to substantiate the effects of AI on student learning outcomes.

## Systematic Review of AI in Mathematics Education

AI's role in mathematics education has significantly evolved over the past decade, reflecting technological and changing educational approaches (Kaushik et al., 2021). Early studies, such as the bibliometric mapping analysis by Hwang and Tu (2021), examined trends in AI applications from 2000 to 2021. The review explored ITS, adaptive learning environments, and AI's impact on problem-solving and conceptual understanding. The authors highlighted the growing use of AI-powered personalized learning and real-time feedback in mathematics education, showing how these tools have shaped individualized learning experiences. Despite these advancements, they emphasized the need for more empirical studies on AI's effectiveness in improving student outcomes, particularly in mathematics.

Building on this foundation, Mohamed et al. (2022) reviewed 20 studies published between 2017 and 2021, focusing on a range of AI technologies, including robotics, teachable agents, and autonomous systems. Their analysis showed how these tools improved problem-solving skills, student engagement, and conceptual understanding. They stressed the transformative potential of AI in personalizing instruction and providing real-time assessment feedback. However, they also noted a lack of qualitative research capturing teachers' and students' perspectives on AI integration, highlighting a key gap in the studies.

Further expanding the scope of research, Opesemowo and Adewuyi (2024) reviewed ten studies published between 2015 and 2023, focusing on AI's impact in the context of the Fourth Industrial Revolution (4IR). Their review highlighted how AI-supported personalized learning and real-time feedback systems could address educational disparities and prepare learners for the evolving demands of the 4IR. The review explored tools like ChatGPT and other AI technologies, which enhance mathematics instruction by providing adaptive learning experiences tailored to individual student needs. However, the authors also emphasized the need for ethical considerations in AI deployment, particularly regarding student data protection and equitable access to AI-assisted learning resources.

While these reviews have advanced the understanding of AI's role in mathematics education, each has limitations. For example, Hwang and Tu (2021) primarily focused on bibliometric trends, offering valuable insights into research patterns but lacking detailed analysis of empirical findings. Mohamed et al. (2022) relied heavily on quantitative data, with limited input from educators and students regarding their experiences with AI integration. Similarly, Opesemowo and Adewuyi (2024) used databases like Scopus and Google Scholar, which may not have captured all relevant studies.

## Purpose Statement

This systematic literature review aimed to identify and evaluate the AI technologies currently used as interventions in mathematics classrooms. The review specifically assessed the effectiveness of these interventions in improving student learning outcomes, such as academic performance, engagement, and personalization. Additionally, the study explored teachers' and students' insights and experiences regarding AI's integration into educational settings. By systematically analyzing the impact of AI across diverse educational environments, the review sought to identify which AI interventions were most effective in enhancing mathematical understanding, engagement, and overall student performance.

## Research Questions

To guide this exploration, the review is structured around two research questions:

- RQ1: What types of artificial intelligence-based interventions have been implemented in mathematics education?
- RQ2: What impact have artificial intelligence-based interventions had on student learning outcomes including academic performance, engagement, and personalization, as observed across the studies in mathematics education?

## Methodology

### Literature Review Protocol

The review was guided by a structured protocol, ensuring consistency and transparency. To ensure a focused scope and address gaps in the literature, the protocol adhered to Preferred Reporting Items for Systematic Reviews and Meta-Analyses (PRISMA) guidelines, which ensured methodological transparency and reproducibility throughout the review process (Page et al., 2021). Multiple academic databases were searched to capture a wide range of peer-reviewed publications, which strengthened the reliability of the findings (Paez, 2017). Consistency was maintained by carefully documenting the inclusion and exclusion criteria, aligned with PRISMA standards. Figure 1 presents a detailed summary of the PRISMA process.

The protocol was a living document, updated regularly as recommended by Bandara and Syed (2023). This iterative approach minimized the risk of bias, improving the quality of the review findings. Including both quantitative and qualitative studies allowed for a comprehensive analysis, capturing measurable outcomes and feedback from teachers and students on the effectiveness of AI interventions (Zheng et al., 2021). This multifaceted strategy aligned with best practices for systematic reviews, ensuring methodological rigor and transparency in assessing the impact of AI technologies in mathematics education (Pigott & Polanin, 2019).

### Databases

The current review addresses limitations identified in prior studies using a broader range of databases, including APA PsycInfo, Education Source, ERIC (via EBSCOhost and ProQuest), Professional Development Collection, Academic Search Premier (all via EBSCOhost), ScienceDirect, Cochrane Library (Wiley), LearnTechLib, ACM Digital Library, IEEE Xplore, SciTech Premium Collection, and Social Science Premium Collection.

This search strategy provides a comprehensive view of AI's role in mathematics education, as advocated by Xiao and Watson (2017). By incorporating studies published between 2020 and 2024, the review captures the latest publications about AI applications, including the use of large language models, adaptive platforms, and AI-enhanced assessments to improve mathematics learning outcomes. The search strategy was developed iteratively, using relevant keywords to identify studies that addressed AI interventions in mathematics education (Page et al., 2021). Predefined inclusion and exclusion criteria ensured the selection of high-quality studies directly relevant to the research questions (Foo et al., 2021).

## Search Criteria

The search strategy employed keywords commonly used to describe AI, mathematics education, and educational interventions in research. These keywords led to the search query of:

("artificial intelligence" OR "AI" OR "a.i." OR "machine learning" OR "intelligent tutoring system") AND ("mathematics" OR "math" OR "math education" OR "mathematics education") AND ("classroom" OR "intervention" OR "instruction").

Database-specific Filters

- APA PsycInfo, ProQuest Education Database, Education Source (EBSCOhost), ERIC (EBSCOhost): Filters were set to include only peer-reviewed journal articles available in full text, published in English, and dated from 2020 onward.
- Cochrane Library: The search in Cochrane Library covered multiple content types, including reviews, protocols, trials, clinical answers, editorials, and special collections. Keywords were applied to titles, abstracts, and keywords for studies from 2020 onward.
- ScienceDirect and LearnTechLib: Searches were conducted to locate full-text, peer-reviewed articles published from 2020 to the present using the keyword query.
- ACM Digital Library and IEEE Xplore: Searches were performed using abstracts and keywords, focusing on articles published from 2020 onward. This filtering by abstracts and keywords helped refine the search results, reducing the retrieval of irrelevant studies where key terms appeared too frequently in the full text.

## Inclusion and Exclusion Criteria

Inclusion criteria required studies to focus on AI interventions in classroom settings, covering from elementary to post-secondary educational levels. Additionally, the selected studies needed to examine the impact of AI on student learning outcomes, engagement, or teaching practices, with some studies including teacher and student perspectives on AI integration. Studies provided detailed methodology sections and key findings were included to ensure the reliability and depth of the analysis. Furthermore, only articles published in English were included to maintain accuracy in data interpretation. Studies were excluded if they did not involve AI interventions in mathematics education or lacked relevance to classroom applications. Research published before 2020 was also excluded to align with the review's emphasis on recent technological advancements.

## Article Selection Process

The article selection process involved a multi-step screening across identified databases, starting with 426 articles. Based on titles and abstracts, initial screening excluded studies which did not meet the inclusion criteria or lacked empirical data on educational outcomes. Duplicates were also removed. The remaining articles underwent a full-text review to ensure methodological rigor and relevance. Ultimately, 14 articles were selected. This process followed PRISMA 2020 guidelines (Page et al., 2021) to ensure transparency and consistency.

## Coding

Selected articles were systematically coded according to key dimensions to provide a comprehensive understanding of AI interventions in mathematics education with a description of each header below:

1. *Author(s)*: The researchers responsible for conducting the study.
2. *Grade Level*: The educational levels targeted by the AI interventions ranged from elementary to post-secondary.
3. *Location*: Geographic regions where the studies were conducted.
4. *Objective*: The primary goal(s) of each study.
5. *Methodology*: The research designs employed in the reviewed studies included experimental, quasi-experimental, mixed-methods, case studies, ethnographic fieldwork, and qualitative case study approaches.
6. *Intervention Duration*: The total length of the AI intervention (e.g., 45 minutes, six weeks). Frequency and session details are included when available.

7. *AI Tools*: The specific AI-based interventions used in each study.
8. *Results*: An overview of each study's key outcomes and findings

This structured approach allowed for a detailed synthesis of how different AI tools are used in mathematics classrooms and their effectiveness in enhancing learning outcomes. By focusing on these aspects, the coding process in this review captures the essential elements of AI integration in education. To ensure consistency and comparability across the reviewed studies, all reported effect sizes were converted to a common metric, Cohen's d, whenever possible. This conversion allowed for standardized comparisons of the magnitude of effects across interventions, facilitating a more cohesive analysis of the study's findings. Cohen's d values were categorized as small ($d$ = 0.2), medium ($d$ = 0.5), or large ($d$ = 0.8) based on conventional guidelines (Cohen, 1988), enabling a clear interpretation of the impact of each intervention.

## Results

The review analyzed 14 empirical studies on AI interventions in mathematics education, spanning elementary to post-secondary levels. These studies offered insights into research designs, educational settings, methodologies, and learner characteristics. Table 1 summarizes key study details, including grade level, locations, objectives, methods, AI tools, duration, and results. The studies occurred in diverse geographic regions, providing a global perspective on AI integration in mathematics education. Methodologies included experimental, quasi-experimental, mixed methods, and case studies.

### RQ1: What types of artificial intelligence-based interventions are currently being implemented in mathematics education?

The reviewed studies explored a range of AI-based interventions in mathematics education, with some types more prominently featured than others. ITS emerged as the most studied intervention in 6 out of 14 studies. Shih et al. (2023) used a dialogue-based ITS to assist students in learning fractions, while Del Olmo-Muñoz et al. (2022) implemented HINTS to improve arithmetic problem-solving. Bush (2021) applied the WMAL system to enhance fraction understanding, and Kooken et al. (2021) validated the MathSpring ITS, which fostered self-regulation and problem-solving skills. Park (2023) used the ASSISTments platform, which, although adaptive, also functions as an ITS by identifying unproductive learning patterns and offering targeted feedback. Shin (2022) investigated prospective teachers' use of ITS tools like MATHia and Khan Academy, commonly applied in mathematics education, to provide individualized, step-by-step guidance and support.

Adaptive learning platforms, which adjust content dynamically based on student performance, were featured in 3 out of 14 studies. Annuš & Kmet' (2024) explored "Learn with ME," a platform for personalized math instruction for elementary students, resulting in significant performance improvements. Phillips et al. (2020) studied ALEKS, a well-known adaptive platform, to supplement algebra instruction. Shin (2022) also explored the integration of ALEKS into lesson planning, focusing on how it helped address individual learning gaps through real-time student assessments.

AI-powered educational apps were examined in 3 out of 14 studies, offering innovative, interactive ways for students to engage with mathematical concepts. Capinding (2023) reported Photomath, an app allowing students to solve problems by taking pictures of equations, significantly improved pre-calculus performance. Wardat et al. (2023) explored ChatGPT as an AI-based tool for mathematics education, finding while it was effective for basic problem-solving, it struggled with more complex concepts like geometry. Cohen et al. (2021) studied an AI-driven recommender system used collaborative filtering algorithms to personalize learning experiences, comparing its effectiveness to tasks assigned by human teachers.

Beyond these tools, other AI interventions were explored in specific contexts. Smart Virtual Reality (VR), used in 1 study, was investigated by King et al. (2024) to train teacher candidates in effective questioning strategies for mathematics instruction, with AI-driven feedback significantly improving pedagogical skills. AI engines for arithmetic practice were featured in Sperling et al. (2022), which examined their use in Swedish primary schools. Although the AI engine personalized learning tasks, it required frequent teacher intervention. Lastly, Egara and Mosimege (2024) studied the integration of ChatGPT into mathematics instruction, noting its potential to increase student engagement and comprehension.

### RQ2: To what extent have artificial intelligence-based interventions improved student learning outcomes within mathematics education?

AI-powered educational apps and adaptive learning platforms have demonstrated substantial improvements in student outcomes with large effect sizes. Capinding (2023) reported that the AI-powered app Photomath led to a 59.4% improvement in pre-calculus performance, with a large effect size (*d* = 1.37), enhancing students' problem-solving abilities and boosting their confidence in mathematics over six weeks. Annuš and Kmet' (2024) similarly found significant improvements with the adaptive platform "Learn with ME," resulting in a performance gain among elementary students *(d* = 0.49) over nine months, with 82.3% of students reporting higher engagement. Del Olmo-Muñoz et al. (2022) noted a 26.7% improvement in arithmetic problem-solving proficiency using the HINTS system, achieving a moderate effect size (*d* = 0.58) after six 45-minute sessions, illustrating how AI interventions can have a notable impact even with moderate effect sizes. These results underscore AI's capacity for delivering significant gains when applied in structured environments tailored to meet students' unique needs.

Several AI interventions showed moderate improvements in learning outcomes, demonstrating effectiveness in specific contexts. For example, Bush (2021) reported the Woot Math Adaptive Learning (WMAL) system improved fraction understanding in 4th and 5th graders, with an effect size of d = 0.39 over 20 instructional days. Shin (2022) also demonstrated that prospective teachers who integrated AI-based tools like Assessment and LEarning in Knowledge Spaces (ALEKS) exhibited moderate improvements in technological pedagogical content knowledge (TPACK), with an effect size of d = 0.50, emphasizing thoughtful technology integration. Additionally, Kooken et al. (2021) validated the Cyclical Self-Regulated Learning (SRL) Simulation Model in the MathSpring ITS, finding a small-to-medium effect size *(d* = 0.26) in mastery scores across countries over a five-week program. These studies highlight that while AI tools may not yield large-scale impacts universally, they are effective in enhancing learning outcomes and engagement when implemented strategically and coupled with thoughtful pedagogical approaches.

Some AI interventions produced limited or context-specific improvements, emphasizing that human mediation and support remain essential. Shih et al. (2023) found that sixth graders using a dialogue based ITS for fractions showed a moderate effect (*d* = 0.59) over a two-hour intervention; however, the short duration limited its potential for producing more sustained learning gains and deeper conceptual understanding. Similarly, Cohen et al. (2021) reported that teacher-assigned tasks in a mathematics applet outperformed machine-assigned ones among 4th and 5th graders, with a small effect size (*d* = 0.13) after two 90-minute sessions, demonstrating the need for personalized human support. Egara and Mosimege (2024) explored ChatGPT's use in Nigerian secondary schools, finding increased student engagement, yet only 17% of teachers were aware of its potential, indicating a need for further professional development. Similarly, Sperling et al. (2022) evaluated an AI engine in Swedish primary schools, noting its personalization capabilities but requiring frequent teacher intervention for two six-week sessions, highlighting the limitations of fully automated systems where teacher support remains essential. King et al. (2024) studied AI-enhanced VR for training teacher candidates in questioning strategies over a four-week period; although improvements were observed, the small sample size (N=2) and unspecified effect size limited generalizability. Wardat et al. (2023) assessed ChatGPT's global use and found it effective for basic problem-solving but less so for complex mathematics like geometry, underscoring the ongoing need for teacher involvement and professional development. Collectively, these findings illustrate that while AI shows promise, broader impacts are limited without human support, emphasizing the importance of context, duration, and teacher involvement to optimize effectiveness.

## Discussion

AI in mathematics education provides valuable opportunities to enhance teaching methods and improve student engagement. Hwang (2022) and Dabingaya (2022) demonstrated that AI-driven personalized learning environments enhance student achievement in mathematics by adapting instruction to individual learning paces. This personalization is especially beneficial in mathematics, where learners often struggle with traditional approaches. Specific AI tools, such as ITS and AI-based calculators, were shown to provide real-time feedback and targeted support, helping students improve their problem-solving skills. These tools can be used to design more effective and accessible mathematics education, especially for students with different levels of mathematical proficiency (Van Vaerenbergh & Pérez-Suay, 2021).

The analysis of differences and patterns in effect sizes, grade levels, and intervention durations reveals several key insights into the effectiveness of AI interventions in mathematics education. Studies with larger effect sizes, such as those by Capinding (2023) and Del Olmo-Muñoz et al. (2022), often involved older students (high school and upper elementary levels) and targeted specific, complex mathematical skills like pre-calculus and arithmetic problem-solving. These interventions typically spanned several weeks or even months, indicating that extended durations allow students to benefit from the adaptive and personalized features of AI tools. The pattern suggests that when AI interventions are applied to complex subjects over longer periods, they offer students the opportunity to practice, receive feedback, and gradually build mastery, leading to significant improvements in learning outcomes (Myers et al., 2022).

Conversely, smaller effect sizes were noted in studies focusing on younger students or shorter interventions. For example, Cohen et al. (2021) and Shih et al. (2023) found only modest gains when AI tools were used briefly or with elementary students, indicating the impact of AI might be limited without sustained engagement. This suggests that for AI to be effective at lower grade levels or in foundational mathematics, it must be integrated consistently over time to foster lasting improvements. The patterns highlight the importance of aligning the intervention type, duration, and grade level; specifically, longer and more intensive use of AI tools may be necessary to see substantial outcomes, particularly when foundational skills are being developed (Dietrichson et al., 2020). While ITS tools show promise at higher grade levels due to their capacity for deep, personalized instruction, adaptive learning platforms appear more versatile across grades but may need longer durations to achieve notable impacts, underscoring the need for strategic implementation in different educational contexts (Alwadei et al., 2020)

As AI continues to be integrated into mathematics education, further research should explore its long-term effects on student outcomes, including how it impacts equity and accessibility. Hwang and Tu (2021) noted the need for studies at different educational levels to assess AI's broader effects, particularly in mathematics, where diverse learning needs must be addressed. Research should also focus on whether AI interventions reduce or widen achievement gaps in mathematics, especially among underrepresented groups.

There are also ethical considerations related to AI use in mathematics education. Privacy, data security, bias, and fairness should be central to future research. AI systems rely on vast amounts of student data, raising concerns about compliance with privacy regulations, such as FERPA in the U.S. and GDPR in Europe. Reiss (2021) stressed the importance of data security measures, including encryption and transparent data handling, to build trust in AI-driven educational tools. Additionally, if left unchecked, bias in AI algorithms could perpetuate inequalities in mathematics education. Baker and Hawn (2021) argued for fairness-aware algorithms and representative datasets to ensure AI tools are equitable for all students, regardless of their background. Continuous monitoring and updates are needed to address biases as AI technologies evolve.

AI systems must be transparent in their decision-making processes, particularly in educational settings where they can influence grades or learning outcomes. Holmes et al. (2021) highlight the importance of transparency in AI tools, ensuring students, teachers, and parents understand how decisions are made. In mathematics education, where AI tools provide personalized learning paths, transparency is essential for fostering trust and ensuring the tools are used appropriately. Accountability protocols must also be established to address errors or unintended consequences, with clearly defined roles for educators, administrators, and developers to prevent misuse of AI systems (Pijls & Dekker, 2011).

While AI has the potential to enhance personalized learning in mathematics, it should support, not replace, human interaction. Teachers bring essential qualities, such as empathy and social engagement, AI cannot replicate. Vasquez et al. (2024) argued AI's primary role in the classroom should be to automate routine tasks and provide data-driven insights, allowing teachers to focus on critical thinking and problem-solving in mathematics. The relationship between teachers and students remains central to effective education, as human interaction fosters the development of emotional intelligence and ethical reasoning, skills crucial to mathematics and broader learning goals (Pila, 2023).

## Limitations

This review encountered several limitations. Variations in methodology, educational levels, and AI technologies across the studies made data synthesis challenging (Visic, 2022). Moreover, the rapid pace of AI development means some earlier studies may no longer reflect the capabilities of newer technologies. These factors limit the generalizability of the findings and necessitate careful interpretation, especially when evaluating AI's impact on mathematics education. Additionally, cultural differences and regional variations were not consistently addressed, further limiting the applicability of the findings to diverse educational systems.

## Conclusion

This literature review demonstrated the positive impact of AI on mathematics education, particularly through tools such as ITS, adaptive learning platforms, and AI-powered educational apps. These technologies showed promise in improving academic performance and student engagement across educational contexts. However, integrating AI into mathematics curricula requires careful consideration of pedagogical practices and ethical concerns. The U.S. Department of Education’s emphasis on AI underscores the need to align AI tools with educational objectives while preserving the essential human aspects of teaching (Ifenthaler & Schumacher, 2023). Although AI interventions show promise at secondary and post-secondary levels, further research is needed to assess their impact on primary education (Salas-Pilco, 2020). Early intervention in mathematics education could be crucial in closing achievement gaps and supporting foundational skills in younger learners (Yazicioğlu, T., & Akdal, 2023).

A discernible pattern emerged when examining the relationship between the duration of AI-based interventions and their associated effect sizes in the literature. Extended interventions, spanning several weeks or months, tend to produce larger effect sizes compared to short-term implementations. Programs like "Learn with ME" (Annuš & Kmet', 2024) and Photomath (Capinding, 2023) exemplify this, showing substantial gains in both student performance and engagement when sustained over time (Saundarajan et al., 2020). Conversely, brief interventions, such as those assessed by Shih et al. (2023) and Cohen et al. (2021), frequently resulted in smaller effect sizes, underscoring the necessity for prolonged engagement with AI tools to achieve significant educational outcomes (Foster, 2023). This pattern underscores that while AI interventions can yield immediate benefits, their maximum efficacy is realized through continuous and long-term integration within educational contexts. These insights are consistent with broader pedagogical research, which highlights the critical role of duration in fostering instructional efficacy and skill development (de Boer et al., 2014). Accordingly, future research should investigate the optimal duration and frequency of AI interventions to maximize their effectiveness, particularly in early education settings, where foundational skills are cultivated (Suggate, 2016).

# CHAPTER THREE: MATHBOT IN MOTION: EVALUATING THE EFFICACY OF AI-POWERED TOOLS FOR ENHANCING FRACTION COMPREHENSION AND SITUATIONAL INTEREST IN ELEMENTARY MATHEMATICS

The chapter titled "Mathbot in Motion: Evaluating the Efficacy of AI-Powered Tools for Enhancing Fraction Comprehension and Situational Interest in Elementary Mathematics" will be submitted for publication in the refereed journal title, *The Journal of Education Sciences: Technology Enhanced Education.*

Abstract

This study examined the effectiveness of Mathbot, an AI-powered educational tool, for enhancing fraction comprehension and situational interest among elementary students with mathematical learning challenges. The study compared outcomes between students using Mathbot and those receiving traditional instruction through a quasi-experimental pretest-posttest design. While both groups showed improvements in fraction comprehension, the traditional instruction group demonstrated greater gains. Additionally, no significant differences were found in situational interest between the two groups, suggesting AI tools like Mathbot may need further refinement to boost engagement.
Keywords: artificial intelligence, mathematics, AI in education, fraction comprehension, situational interest, personalized learning, elementary mathematics, educational technology, machine learning, adaptive learning tools, AI-powered education, mathematics instruction

# Introduction

Fractions are a fundamental but challenging area of elementary mathematics, with concepts like part-whole relationships, equivalence, and proportionality often confusing learners and impeding mastery (Getenet & Callingham, 2019). For students with mathematical learning challenges (MLC), these difficulties are often exacerbated by traditional instructional methods, which may not address their diverse needs. Despite years of research and pedagogical advancements, nationwide assessments indicate only 36% of 4th graders and 26% of 8th graders are proficient in mathematics, with even fewer demonstrating proficiency in fraction-based problems (National Center for Education Statistics, 2022a, 2022b, 2022c). These statistics underscore a persistent gap in mathematical understanding, especially in foundational areas like fractions, which are crucial for advanced learning.

Addressing these challenges requires innovative strategies that go beyond traditional teaching methods. Research shows interventions using conceptual and linguistic scaffolds, such as manipulatives like elastic strips, enhance students' understanding of fractions by providing tangible representations of abstract concepts (Harvey, 2012; Prediger & Wessel, 2013). While effective, these approaches rely heavily on teacher-directed instruction and may not fully leverage the potential of personalized learning pathways, which can adapt to each student's pace and needs (Lee et al., 2018). Tailored, student-centered methods offer flexibility, adaptability, and real-time feedback to foster deep conceptual understanding.

In this context, the integration of Artificial Intelligence (AI) in education offers a promising solution. AI-powered tools like Mathbot can provide personalized learning experiences by adapting instruction to each student's strengths and learning patterns (Lee et al., 2018). These tools can adjust difficulty levels, offer immediate feedback, and present content in varied formats to accommodate different learning styles (Du Boulay, 2016; Sun et al., 2020).

AI's ability to collect and analyze real-time data enables it to provide insights inform instructional decisions, supporting more effective and equitable educational outcomes (Ifenthaler & Schumacher, 2023). For instance, Mathbot addresses gaps in fraction comprehension by customizing lessons based on each student's progress, offering scaffolding where needed, and gradually increasing complexity as understanding deepens (Reinhold et al., 2020).

Beyond improving comprehension, AI tools like Mathbot can enhance situational interest, short-term, context-specific curiosity drives engagement in learning (Al Omoush & Mehigan, 2023; Garrett et al., 2020). This is particularly important for students with MLC, who may struggle with sustained engagement due to frustration or boredom with traditional methods. By incorporating interactive, adaptive elements, AI tools can provide a more engaging learning experience, making complex concepts like fractions more accessible and enjoyable (Akavova et al., 2023; Hunt et al., 2023).

Despite the growing body of research on AI in education, empirical studies rigorously evaluating its effectiveness in specific content areas, such as fraction comprehension, remain limited. Early studies show promise, but questions remain about best practices for integrating AI tools into classrooms and their long-term impact on learning outcomes (Silva & Janes, 2020). Additionally, AI's potential to foster more equitable learning environments has not been fully explored. Further research is needed to determine how AI can close achievement gaps for students with learning challenges, ensuring all students have the opportunity to succeed in mathematics (Holstein & Doroudi, 2021).

This study seeks to fill this gap by evaluating the effectiveness of Mathbot for enhancing fraction comprehension and situational interest among elementary students with MLC. By focusing on a group of students often underserved by traditional methods, this research aims to provoide insights into how AI-driven personalized learning environments can support more inclusive and effective educational practices. The findings will contribute to the growing literature on AI in education and offer practical recommendations for educators and policymakers about integrating AI tools to promote mathematic learning outcomes.

## Literature Review

### The Need for Innovative Educational Strategies

The current state of mathematics proficiency among U.S. students has been well-documented, with studies pointing to a decline exacerbated by the COVID-19 pandemic, amplifying the call for innovative strategies (National Center for Education Statistics, 2022; Fuchs et al., 2023; Kuhfeld et al., 2022). Mathematics, foundational to academic success and practical daily numeracy, requires effective methodologies to foster a deep understanding. Research indicates active methodologies are fundamental to academic success (Candela-Muñoz & Rodríguez-Gámez, 2023), and teachers' mathematical knowledge is linked to student achievement gains. However, recent studies suggest this relationship may be complex and not always directly related to student performance (Hill et al., 2005; Sutherland et al., 2022). Problem-oriented teaching methods have proven effective in improving mathematics achievement and reducing disparities between different student demographics (Tu et al., 2018).

Within this context, game-based learning, aligned with a Universal Design for Learning (UDL) framework, emerges as a potent pedagogical strategy. Hunt et al. (2023) demonstrated digital games, which could be powered by AI, enhanced student engagement, fraction knowledge, and STEM interest. These games provided personalized learning pathways which cater to individual needs, offering the potential to bridge gaps in understanding, particularly for students with learning challenges (Jian, 2023; Pappas & Drigas, 2016). This approach not only increases interest and accessibility, but also holds promise for addressing equity issues by ensuring all students, regardless of ability, can engage meaningfully with complex mathematical concepts.

## AI in Education

AI technologies, such as machine learning and natural language processing, enable the creation of adaptable and customized content catering to students' diverse learning profiles (Zhao & Li, 2020). This adaptability has been shown to improve student outcomes, increase motivation, and heighten situational interest (Hui et al., 2023; Rajeswari & Purushothaman, 2023). A systematic analysis by Mohamed et al. (2022) examined 20 studies published between 2017 and 2021. Findings from the study demonstrated AI can enhance engagement, conceptual understanding, and problem-solving skills. AI can provide personalized learning experiences, as evidenced by promising results across numerous studies (e.g., Ashwini et al., 2023; Chen et al., 2020; Liu et al., 2022).

Opesemowo & Adewuyi (2024) noted AI-driven personalized learning has the potential to transform mathematics education by providing real-time feedback and individualized learning pathways. This ensures students receive the specific support they need, which is essential for addressing educational disparities. AI’s ability to cater to learner variability aligns with broader efforts to ensure educational equity and access, particularly in mathematics, which is foundational to many academic disciplines.

AI's adaptability to individual learning preferences is especially valuable in mathematics education. Studies show AI-powered systems can assess a student's proficiency and tailor learning materials to their level, offering immediate, personalized feedback, enhancing the learning experience (Pappas & Drigas, 2016; Seo et al., 2021). AI technologies have been applied in various educational contexts, from primary school mathematics (Liu et al., 2022) to college-level instruction (Ma, 2023). This research demonstrated AI can effectively support diverse educational needs across different levels.

AI technologies, such as expert systems, intelligent tutors, and adaptive assessments, offer nuanced insights into a student's learning process by analyzing data and providing personalized feedback. These insights can lead to substantial improvements in student learning outcomes (Kochmar et al., 2020). This is particularly beneficial in subjects like mathematics, where cumulative learning and mastery of foundational concepts are crucial for progression (Chang & Lu, 2019). Opesemowo & Adewuyi (2024) further emphasized AI technologies provide personalized learning trajectories and interventions, allowing educators to identify and address specific student needs early. This early intervention is essential for ensuring no student falls behind, especially in subjects like mathematics where concepts build on one another.

### Challenges and Ethical Considerations in AI Education

The integration of AI into education is not without challenges. For AI to be truly effective, it must enhance, rather than replace, teacher-student interactions (Vasquez et al., 2024). AI systems must be designed to complement the pedagogical goals of educators and adapt to specific classroom contexts (Seo et al., 2021; Tapalova & Zhiyenbayeva, 2022). Additionally, the use of AI in educational settings raises concerns regarding equity and access. It is crucial to ensure all students, regardless of socio-economic background, have equal opportunities to benefit from personalized learning experiences (Maghsudi et al., 2021; Popenici & Kerr, 2017).

The U.S. Department of Education (2023) stresses the importance of maintaining human oversight when deploying AI systems, ensuring AI supports rather than replaces the human elements of teaching. This aligns with the need to prevent algorithmic bias and ensure transparency in AI-powered educational tools. Furthermore, policymakers are urged to establish clear guidelines to protect student privacy and prevent potential misuse of AI-generated data.

Enhancing Situational Interest through AI

Situational interest, which is crucial for sustained student motivation, is significantly influenced by the learning environment (Bolkan & Griffin, 2018). Hidi and Renninger (2006) defined situational interest as the initial curiosity sparked by engaging external stimuli, which can evolve into deeper, more persistent interest. AI-enhanced learning environments, featuring tools such as intelligent tutoring systems, gamified modules, and dynamic simulations, provide such stimuli. These systems create immersive, adaptive scenarios encouraging deeper interaction with educational material. By responding to individual student needs, AI can deliver tailored learning experiences fostering engagement and making learning more enjoyable (Pappas & Drigas, 2016).

Daghestani et al. (2020) demonstrated how gamification, combined with educational data mining techniques, enhanced student engagement by personalizing learning pathways. By adjusting game elements, such as challenges, leaderboards, and rewards, AI-powered gamified systems created dynamic learning experiences nurturing situational interest in subjects like mathematics and science. This adaptability ensured content aligned with each student's pace and proficiency, making complex concepts more approachable. Additionally, these AI-driven systems helped close achievement gaps by addressing the diverse needs of learners, particularly those with learning challenges. By tailoring content to individual needs, AI supports equitable educational experiences, ensuring personalized instruction meets both engagement and learning objectives (Holstein & Doroudi, 2021; Hunt et al., 2023). As Hui et al. (2023) emphasized, personalization is key to maintaining situational interest, especially in challenging subjects where students might otherwise lose motivation.

## Purpose of the Study

This study aimed to evaluate the effectiveness of Mathbot, an AI-powered personalized learning tool for enhancing fraction comprehension and situational interest among elementary students with mathematical learning challenges (MLC). Situational interest, defined as short-term curiosity stimulated by engaging external stimuli, has been shown to significantly improve student engagement and motivation, particularly in mathematics (Wang et al., 2021). For students with MLC, maintaining situational interest is critical for sustaining focus and attention, which fosters deeper comprehension of foundational concepts such as fractions (Lindstetd et al., 2020). Mathbot's ability to deliver adaptive lessons and immediate feedback tailored to individual needs offers more targeted support than traditional instruction (Akavova, 2023). By focusing on both fraction comprehension and situational interest, this study seeks to determine whether AI-driven learning environments can provide more personalized and effective educational experiences.

## Theoretical Framework

This study is grounded in two key educational theories: Seymour Papert's constructionism and B.F. Skinner's reinforcement learning. These frameworks complement each other in explaining how technology can enhance both knowledge construction and the reinforcement of positive learning behaviors. Papert emphasizes active learning through interaction and reflection, while Skinner focuses on shaping behavior through reinforcement. Together, these theories provide a solid foundation for designing and evaluating AI-driven learning tools like Mathbot.

Papert's Constructionism highlights the importance of learning through hands-on experiences and active engagement. In this study, Mathbot operationalized these principles by allowing students to engage directly with fraction problems through interactive problem-solving tasks. The system provided immediate feedback, enabling students to reflect on their answers and adjust their strategies in real time. This active, iterative process of solving problems, receiving feedback, and refining their understanding aligned with Papert's (1985) view learners constructed knowledge most effectively when they were actively involved in their learning. Rather than passively receiving information, students were encouraged to develop their understanding through trial, reflection, and adjustment, embodying the core tenets of constructionism. Additionally, group discussions after Mathbot sessions provided opportunities for reflection and peer-to-peer learning, further reinforcing Papert's ideas about the importance of interaction and reflection in deepening understanding.

Skinner's Reinforcement Learning Theory emphasizes the role of immediate feedback in shaping learning behaviors. In Mathbot, students received immediate positive reinforcement for correct answers, which encouraged the repetition of successful problem-solving strategies. When students made mistakes, the system provided additional scaffolding and prompted them to review and practice further, guiding them toward mastery. This feedback loop of reinforcing correct responses while offering support for incorrect ones reflects Skinner's belief that timely reinforcement strengthens desired behaviors and promotes effective learning (Skinner, 1953). By immediately addressing misconceptions and providing opportunities for further practice, Mathbot supports behavioral shaping in real time, helping students progressively develop their comprehension of fractions.

# Methodology

This study was conducted with the approval of the Institutional Review Board (IRB) at the University of Central Florida, ensuring adherence to ethical research standards. IRB approval was granted on 1/25/24 with approval # FWA00000351, IRB00001138, IRB00012110.

## Research Design

The study used a quasi-experimental pretest-posttest design with non-randomized groups. One group received Mathbot's AI-powered personalized learning, while the other followed traditional business-as-usual (BAU) teaching methods. Both groups were assessed using the same pre-test and post-test measures, which allowed for a clear comparison of changes in fraction comprehension and situational interest over time.

The quasi-experimental design was chosen to preserve the natural classroom dynamics, allowing the study to be conducted without disrupting the existing structure of the school environment. As Misquitta (2011) pointed out, this design is ideal when random assignment is not feasible due to practical constraints, such as maintaining natural classroom groupings. By using intact classrooms rather than randomly assigning students to groups, the study achieves a high level of ecological validity, meaning the findings are more likely to generalize to typical classroom settings (Chezan et al., 2020).

The pretest-posttest structure tracked student performance and interest changes over time. The pre-test provided baseline data on students' fraction knowledge and situational interest, allowing for an initial comparison between groups, while the post-test revealed how each group's comprehension and interest evolved after the interventions. This design allowed the study to capture both immediate learning outcomes and changes in student interest, providing deeper insights into the effectiveness of the intervention.

Although random assignment was not used, the quasi-experimental approach remains a robust method for investigating the effectiveness of different teaching strategies (Bärnighausen et al., 2017). By comparing the AI-powered Mathbot instruction to traditional teaching, this design provides insights into how these methods affect student learning in an actual classroom setting. The combination of both pre-test and post-test measures further strengthens the study's ability to draw valid conclusions about the effectiveness of Mathbot, particularly in improving both learning outcomes and situational interest.

Research Questions and Hypotheses

This study addressed two key research questions:

Research Question 1 (RQ1): Is there a significant difference in the change in understanding of fractions between students with mathematical learning challenges using the Mathbot and those in the Business-as-Usual condition?

1. Null Hypothesis ($H_0$): There is no significant difference in the change in understanding of fractions between students with mathematical learning challenges using the Mathbot and those in the Business-as-Usual condition.
2. Alternative Hypothesis ($H_1$): There is a significant difference in the change in understanding of fractions between students with mathematical learning challenges using the Mathbot and those in the Business-as-Usual condition.

Research Question 2 (RQ2): Is there a significant difference in the change in self-reported situational interest between students using the Mathbot and those in the Business-as-Usual condition?

1. Null Hypothesis ($H_0$): There is no significant difference in the change in self-reported situational interest between students using the Mathbot and those in the Business-as-Usual condition.
2. Alternative Hypothesis ($H_1$): There is a significant difference in the change in self-reported situational interest between students using the Mathbot and those in the Business-as-Usual condition.

These research questions examine the dual impact of the AI-powered Mathbot on students with mathematical learning challenges, proficiency, and situational interest, probing not only quantitative advancements in fraction comprehension but also potential changes in learners' attitudes toward mathematics.

### Participants and Sampling

#### Inclusion and Exclusion Criteria

Students were included in the study if they were identified as having Mathematical Learning Challenges (MLC), defined as persistent difficulties in understanding and applying fraction concepts. Eligibility was determined by performance below proficiency on state standardized tests, a grade of 'C' or lower on fraction-related assignments, or teacher recommendations based on observed struggles with mathematical concepts. Students were excluded if they had demonstrated proficiency in fractions, missed more than 20% of the intervention, or transferred out of the school.

#### Recruitment of Students

Students were recruited from two 4th and 5th-grade classrooms in the southeast United States. Teachers were briefed on the study's methodology. They recommended students meeting the inclusion criteria for MLC. Based on these recommendations, 43 students were selected. Parental consent was obtained through letters outlining the study's purpose, and all required forms were collected before the intervention. Recruitment was completed over two weeks to minimize classroom disruption.

Sampling Method

Convenience sampling was used due to the practical constraints of a pilot study. This method allowed for the quick recruitment of 4th and 5th graders with MLC, ensuring the study reflected real classroom environments. While convenience sampling may limit the generalizability of findings, its ecological validity supports understanding how Mathbot functions in practice (Chezan et al., 2020).

Power Analysis

An a priori power analysis was conducted using G*Power (version 3.1.9.6) to determine the necessary sample size for adequate statistical power in this study's design. The study employed a mixed-design ANOVA (repeated measures within-between subjects factors), comparing the Mathbot intervention group and the Business-as-Usual (BAU) control group across two time points (pre-test and post-test).

The effect size measure used for the power analysis was Cohen’s f, set at 0.25, representing a medium effect size based on Cohen's (1988) conventions for educational research. This corresponds to a partial eta-squared ($\eta^2$) of approximately 0.06. This effect size was chosen based on previous studies demonstrating medium effects from short-duration, technology-enhanced interventions (Jupri et al., 2015; Kraft, 2020). Using an alpha level of 0.05 and a desired power of 0.80, the power analysis indicated that a total sample size of 24 participants (12 in each group) would be required to detect a medium effect size for the interaction between Group and Time.To account for potential attrition (estimated at 20%), a target recruitment of 29 participants was set.

## Instruments

### Fraction Comparison Assessment (FCA)

The Fraction Comparison Assessment (FCA) was developed to evaluate students' proficiency in understanding and comparing fractions, a key competency in elementary mathematics. Using a Delphi process, two experts, Dr. Jessica Hunt (special education fractions, NC State University) and Dr. Sarah Bush (general education fractions, University of Central Florida), contributed to five rounds of revisions, ensuring consensus on the assessment's content and structure. This process aligned the FCA with educational standards and ensured it effectively measured the targeted fraction concepts (Landeta, 2006).

The FCA, as shown in Appendix C, consists of 10 multiple-choice questions in a paper-and-pencil format. These questions range in difficulty and cover a variety of fraction concepts, including equivalence, comparison of magnitudes, and interpretation of visual fraction models. The FCA aligns with the national Common Core State Standards for 4th and 5th-grade students. To minimize the likelihood of memorization and control for potential test-retest effects, the order of the questions was rearranged between the pre-test and post-test. While the same concepts were assessed in both tests, the changes in question order were intended to reduce the risk of students simply remembering the questions. However, it should be acknowledged some potential for memorization remains, which is a limitation of using multiple-choice formats over a relatively small number of questions.

Each question in the FCA is scored as correct or incorrect, and the total number of correct answers provides a percentage score ranging from 0% (no correct answers) to 100% (all correct answers). While these scores indicate varying levels of proficiency, it should be noted a high score does not necessarily imply "full mastery" of the concepts, as multiple-choice assessments can have inherent limitations, such as the possibility of guessing. Instead, the results reflect the students' demonstrated understanding of fraction comparison concepts, with higher scores suggesting a firmer grasp of the material.

## Situational Interest Survey (SIS)

The Situational Interest Survey (SIS), as shown in Appendix D, was adapted from Mitchell’s (1993) original measure of situational interest in mathematics classrooms. It consists of six Likert-scale items rated on a 6-point scale, which assess students’ responses to the Mathbot intervention and traditional instruction (BAU). Lower scores indicate higher situational interest. The six selected items, part of a validated subtest from Mitchell’s original 38-item survey, achieved a Cronbach’s alpha of 0.90, indicating strong reliability. In this study, the recalculated Cronbach’s alpha was 0.83, confirming good internal consistency.

The SIS aligns with the foundational work of Hidi and Renninger (2006) on interest development, as well as subsequent studies by Linnenbrink-Garcia et al. (2010) and Rotgans and Schmidt (2011), which highlight how situational interest can be sparked by engaging external stimuli in the learning environment. By focusing on interest, mainly the curiosity and attention students give to their learning tasks, the SIS provides detailed insights into their motivational responses to the Mathbot intervention. However, while the SIS offers valuable information on students’ interest development, reliance on self-reported data poses a limitation. Self-reported measures are prone to social desirability bias, potentially skewing the accuracy of responses (Tan et al, 2021).

### Classroom and Instructional Setting

The study was conducted in one 4th-grade and one 5th-grade classroom in an inclusive suburban school in the southeast United States. This setting provided an authentic learning environment where the intervention could be implemented without disrupting the natural flow of the school day.

*Classroom Setting*

Both classrooms had typical educational resources, including whiteboards, desks arranged in groups, and access to technology such as tablets and computers. The classrooms followed a structured daily schedule, with a dedicated 40-minute period each day for math instruction. The physical setup of the classrooms allowed for both whole-class instruction and independent learning activities, which was crucial for implementing both the traditional and AI-powered instructional methods.

The inclusive nature of the classrooms ensured students with diverse learning needs, including those identified with MLC, could participate in the study. This diversity in the classroom population reflects the broader challenges of providing effective instruction for students with varying proficiency levels in mathematics, making the findings applicable to real-world educational settings.

*Instructional Setting*

In both the experimental and control groups, fraction-based content was taught during the designated math period. Although the instructional methods varied between the two groups, both received the same mathematical content for the same duration, ensuring consistency in curriculum delivery across the classrooms.

- In the control group**,** instruction followed traditional, teacher-led methods without the integration of AI tools.
- In the experimental group**,** the Mathbot AI-powered tool delivered personalized instruction, adapting to each student's learning pace and interests.

## Intervention

Mathbot is an adaptive, web-based application which customizes fraction problems according to each student's interests and learning pace. The tool integrates various elements, such as sports, music, and hobbies, to create math problems resonating with students' preferences. Through natural language processing (NLP), Mathbot engages students in interactive problem-solving tasks, providing immediate feedback and progressively adjusting the difficulty of the problems. This AI-driven platform promotes mastery by offering personalized hints and scaffolding, allowing students to build upon their existing knowledge and address specific conceptual gaps in their understanding of fractions.

At the core of Mathbot's technology is ChatGPT 4-Turbo, accessed via chatbotkit.com, which enhances the personalization of learning for 4th and 5th-grade students (Mohamed et al., 2022). By emulating a traditional teacher-student interaction, Mathbot creates a dialogue both adaptive and responsive to student inquiries, adjusting explanations and pacing to meet each student's needs. This interaction mirrors the kind of personalized feedback typically provided by human educators, demonstrating the potential of AI to scaffold knowledge acquisition and provide a highly tailored educational experience.

Mathbot was developed to bridge the gap in personalized education, using AI to create customized math learning environments for 4th and 5th graders. Its innovative design weaves personal interests into lessons, enhancing situational interest and motivation. ChatGPT 4-Turbo was selected for its ability to deliver nuanced responses to students' questions, making it a powerful resource for interactive dialogue reflecting teacher-student dynamics. The design methodology, deeply rooted in the belief learning should be personalized and accessible, was rigorously tested to demonstrate the transformative potential of AI in education (Bernacki et al., 2021). Leveraging chatbotkit.com enabled the integration of Mathbot's interactive components quickly and efficiently, ensuring the tool could be utilized promptly for educational purposes.

One of Mathbot's standout features is its ability to instantly incorporate students' interests and hobbies into fraction problems. This customization occurs through an interactive questionnaire where students input their interests at the beginning of the session. These interests are then stored in a personalized chat save file. Using advanced machine learning algorithms and NLP, Mathbot dynamically fetches these interest profiles during each session to generate fraction problems incorporating elements familiar to the student, such as sports statistics, music theory, or relevant cultural references. This real-time customization enhances student interest and aids comprehension by connecting abstract mathematical concepts to meaningful and familiar contexts for the learner.

For example, a student interested in sports might receive fraction problems framed around player statistics or game scenarios. In contrast, another student interested in music might encounter problems involving rhythm and timing. This approach personalizes the learning experience to foster motivation and a deeper understanding of mathematical concepts. Mathbot's backend system, powered by a large language model (ChatGPT 4), continuously processes these personalized elements to create unique and individualized problems. The system's instructional framework is aligned with the student's grade level and is designed to provide a safe, judgment-free environment where students can openly interact with the AI. This adaptive, responsive, learning environment allows students to ask questions, receive tailored feedback, and progress independently.

The backend training process for Mathbot, detailed in Appendix E, demonstrates how the tool's algorithms were developed to tailor educational content for each student. This process was guided by principles of personalized education, ensuring Mathbot could respond to the individual needs of each learner while maintaining a consistent focus on the educational goals of the study.

For a visual representation of Mathbot's interface and its interactive components, refer to Figure 2.

## Lesson Plans

A 5-day lesson plan, presented in Appendix F, was developed to guide the BAU group, with a design paralleling the Mathbot intervention. These plans focused on fraction comparison, each day addressing different aspects of fraction comprehension. These lesson plans were also foundational for training Mathbot, ensuring alignment between the content delivered by the teacher in the BAU condition and the AI-powered personalized instruction provided by Mathbot.

The lesson plans were crafted through a Delphi process involving two general and special education mathematics experts. This consultation ensured the content was pedagogically sound and aligned with educational standards for fractions, promoting a consistent and equitable learning experience across both groups.

Each lesson followed a structured script, which included:

- Introduction: A discussion introducing the day's mathematical concept, such as fraction comparison or equivalence.
- Distributed Practice: Guided practice problems where students worked individually or in groups to apply the concept in different contexts.
- Discussion: Students compared fractions and explained their reasoning, fostering critical thinking and deeper understanding.
- Closing & Check for Understanding: Each lesson ended with an exit ticket or a similar activity to verify student comprehension of the day's material.

Teachers were provided with the lesson plans two weeks before the study to allow sufficient time for review and preparation. This approach ensured instruction in the BAU group adhered closely to the standardized plan, and Mathbot's AI was trained based on the same instructional content. This uniformity across instructional approaches allowed for a fair and reliable comparison between the traditional classroom (BAU) and the AI-supported (Mathbot) teaching methods.

The 5-day lesson plan was designed to achieve specific learning goals for each day (see Table 2).

## Procedures

### Pre-Intervention Activates

### Teacher Training

A comprehensive one-hour Zoom training session was conducted for the classroom teachers involved in both the Mathbot intervention group and the BAU control group two weeks prior to the start of the intervention. The session provided a review of the structured 5-day lesson plan used across both groups ensuring uniformity in objectives and pacing. The training covered detailed instructions on introducing each concept, guiding practice activities, facilitating classroom discussions, and assessing student understanding. Emphasis was placed on administering the pre- and post-intervention assessments, including the Fraction Comparison Assessment (FCA) and the Situational Interest Survey (SIS). Teachers were instructed to introduce these assessments consistently and to refrain from aiding beyond clarifying procedural aspects.

For the Mathbot intervention group, additional training was provided on using the Mathbot tool effectively in the classroom. Teachers were extensively trained on the tool's functionality, including how to navigate its features, troubleshoot common technical issues, and what steps to take in case of any difficulties. Clear instructions were also provided for contacting the researcher if technical issues arose during the intervention. The Zoom training session was recorded, allowing teachers to revisit the content throughout the study to maintain fidelity to the intervention protocol.

### Pre-Assessments and Baseline Setup

Before the intervention began, all students in the Mathbot intervention group and the BAU control group completed two assessments: the FCA and the SIS. These pre-test assessments were designed to establish baseline data for both groups, enabling a comparison of their fraction comprehension and situational interest before the intervention. Both assessments were administered in the classroom under the supervision of the student's respective teachers. Detailed written instructions were provided to the teachers on how to present the assessments, including guidelines for time allocation and permissible clarifications. The assessments were given across all classrooms during the same time period to minimize external variables. Teachers were instructed to refrain from offering any assistance beyond clarifying test procedures. All collected data were securely stored and prepared for subsequent analysis to ensure consistency and accuracy in the study's results.

### Intervention Implementation

The intervention was conducted over five consecutive school days, with each session lasting 40 minutes during the regularly scheduled math period. The Mathbot intervention and BAU control groups were exposed to the same fraction-based content, ensuring lesson objectives and pacing uniformity. The key difference between the two groups lay in the method of instruction: the Mathbot group used an AI-powered tool for personalized learning, while the BAU group followed traditional teacher-led instruction. Both groups followed the same five-day instructional plan, with each day focused on different aspects of fraction comprehension.

### Mathbot AI-Powered Personalized Instruction

The Mathbot classroom utilized an AI-powered personalized learning tool, adapting fraction instruction to meet the individual needs of each student. Mathbot provided real-time feedback and supported students based on their performance, creating a tailored learning path for every participant. This method ensured students received immediate help with misconceptions, allowing for more individualized learning.

Students worked independently using Mathbot to engage with interactive fraction problems aligned with the same fraction content covered in the BAU classroom. Instead of teacher-led instruction, the AI tool guided the learning process, offering step-by-step instructions and feedback as students progressed through the material. Teachers in the Mathbot classroom took on a facilitator role, ensuring students stayed engaged and focused, supporting them with technology navigation, but allowing the AI system to provide the primary instructional guidance.

### BAU Traditional Instruction

In contrast, the BAU classroom relied on a standard mathematics curriculum delivered through direct instruction without the integration of AI-powered tools. The lessons were presented through teacher-led instruction, where the teacher introduced fraction concepts followed by distributed practice, during which students completed exercises from textbooks and worked through examples to reinforce their understanding. While effective for some learners, this approach can overlook the diverse learning needs of students, particularly those struggling with mathematical concepts like fractions (Martin & Hunt, 2022).

The teacher facilitated discussions to support comprehension, where students compared fractions and explained their reasoning. This approach promoted critical thinking and peer interaction, though research suggests traditional methods like these may not effectively maintain situational interest, especially for students facing learning challenges (Boaler, 1996; Lesani et al., 2017). The BAU method lacked the personalized feedback necessary for addressing individual difficulties in real-time, an advantage of AI-based tools. While teachers can help some students immediately, BAU instruction does not provide the consistent, tailored interventions needed to address every student's misconception, which can be crucial for those with mathematical learning difficulties.

### Consistency and Monitoring

To ensure consistency between the two groups, the Mathbot and BAU groups followed the same daily lesson plans, which outlined the fraction topics to be covered and the specific learning objectives for each day. This allowed for a direct comparison of student outcomes between the groups, as the content was consistent across both conditions. Teachers in both groups were given the same instructional timelines and objectives, with the only difference being the mode of instruction.

Throughout the intervention, the research team maintained regular contact with the teachers to monitor the implementation process and ensure fidelity of the instructional approach. Teachers in the Mathbot group were provided with guidelines on managing the classroom during the intervention, including protocols for addressing technical issues. Teachers were encouraged to provide feedback on the implementation's progress, including any classroom engagement issues or challenges in adhering to the lesson plans. This feedback helped ensure consistency in content delivery and instructional pacing. Classroom observations were conducted to ensure teachers in both groups adhered to the lesson plans and the learning environment remained conducive to student interest.

### Handling Disruptions or Technical Issues

Teachers in the Mathbot group were provided with guidelines on managing the classroom during the intervention, including protocols for addressing technical issues. Any technical difficulties encountered with Mathbot were reported directly to the researcher for immediate resolution. In the event of significant technical issues, such as a class-wide system failure or internet outage, students in the Mathbot group would pause AI-based learning for the session and resume it the following day. To make up for the missed session, the schedule would be adjusted to complete all five Mathbot sessions, preserving instructional consistency.

### Fidelity of Implementation

The fidelity of implementation was closely monitored throughout the five-day intervention to ensure the Mathbot and BAU conditions were delivered as intended. For the BAU group, classroom observations were conducted to verify adherence to the lesson plans, and the teacher was asked to document daily activities. These observations ensured instructional time and content aligned with the pre-designed lesson scripts. Additionally, audio recordings captured classroom interactions, verifying the lessons were delivered consistently.

For the Mathbot group, fidelity was monitored through the teacher's documentation and data tracking provided by Mathbot's backend system. The system recorded chat logs between students and the AI, which were reviewed to ensure the personalized learning experience was consistent across participants. The chat logs provided insights into student interactions with the tool, including the number of questions asked, completion of fraction problems, and performance progress. Analytics data, such as the number of interactions and the content adaptation to individual student needs, further confirmed students were engaging with the Mathbot tool as intended.

Any discrepancies or deviations from the planned instructional methods were identified through these monitoring tools and addressed immediately to maintain the integrity of the study.

Combining classroom observations, audio recordings, and data analytics ensured high fidelity to the instructional approaches in both groups.

### Post-Intervention Activities

After completing the five-day intervention, the Mathbot intervention group and the BAU control group participated in post-assessments to measure changes in fraction comprehension and situational interest. The assessments were the same as the pre-intervention tests and were administered consistently to ensure reliability.

### Post-assessment

Teachers were provided instructions to ensure the consistent administration of the post-assessments across both groups. Absentees were allowed to complete the assessments one day after the post-assessment to ensure complete data collection. After collecting the assessments, the data were securely stored and anonymized to protect student confidentiality. Teachers were debriefed on the study and allowed to provide feedback. Any technical or procedural issues encountered during the intervention were noted for future improvements. The collected data were then prepared for statistical analysis to evaluate the intervention's effectiveness in improving fraction comprehension and situational interest.

## Analysis

### Purpose

This analysis evaluates the effectiveness of the Mathbot intervention in enhancing students' comprehension of fractions and situational interest over time. The goal was to determine whether the Mathbot group shows more significant improvements than the BAU group, thereby highlighting how adaptive, AI-powered educational tools can support personalized learning environments.

Research Questions

This analysis addresses the following research questions:

Research Question 1 (RQ1): Is there a significant difference in the change in understanding of fractions between students with mathematical learning challenges using the Mathbot and those in the Business-as-Usual condition?

a) Null Hypothesis ($H_0$): There is no significant difference in the change in understanding of fractions between students with mathematical learning challenges using the Mathbot and those in the Business-as-Usual condition.
b) Alternative Hypothesis ($H_1$): There is a significant difference in the change in understanding of fractions between students with mathematical learning challenges using the Mathbot and those in the Business-as-Usual condition.

Research Question 2 (RQ2): Is there a significant difference in the change in self-reported situational interest between students with mathematical learning challenges using the Mathbot and those in the Business-as-Usual condition?

a. Null Hypothesis ($H_0$): There is no significant difference in the change in self-reported situational interest between students with mathematical learning challenges using the Mathbot and those in the Business-as-Usual condition.
b. Alternative Hypothesis ($H_1$): There is a significant difference in the change in self-reported situational interest between students with mathematical learning challenges using the Mathbot and those in the Business-as-Usual condition.

Variables and Measures

Independent Variables (IV)

1. Condition: Mathbot group (treatment) or BAU group (control).
2. Time: Pre-test and post-test

### Dependent Variables (DV)

1. Fraction Comprehension: Measured by the FCA
2. Situational Interest: Measured by the SIS

### Analytical Approach

Repeated Measures ANOVA (RM-ANOVA) was used to assess the impact of the Mathbot intervention on students' fraction comprehension and situational interest. The analysis evaluated changes within subjects (over time) and between subjects (Mathbot vs. BAU groups), identifying main effects and interaction effects.

### Main Effects

The main effect of time assessed whether there was a significant change in fraction comprehension and situational interest from pre-test to post-test across all participants. The main effect of condition examined differences in these outcomes between the Mathbot and BAU groups, regardless of time.

### Interaction Effects

The interaction effect between time and condition identifies whether changes in comprehension and interest over time varied between the Mathbot and BAU groups. This distinction is crucial as it determines whether Mathbot had a differential impact on learning outcomes, potentially accelerating learning or interest compared to traditional instruction.

### Data Preparation

To maintain data integrity, participants who missed more than 20% of the intervention (i.e., one or more days) or did not complete the pre-assessment or post-assessment were excluded. Outliers were also removed using a Z-score cutoff of $\pm 2$, following standard practices in statistical analysis (Schober et al., 2021). This step was particularly important given the small sample size, as outliers could disproportionately affect the results. After accounting for attrition, which reduced the sample from 43 to 29, and the removal of 7 outliers, the final sample consisted of 22 participants: 13 in the Mathbot group and 9 in the BAU group (see Table 3).

### Descriptive Statistics

Descriptive statistics (means and standard deviations) were calculated for the FCA and SIS scores at pre-test and post-test for both groups (see Table 3 and Table 4).

### Software

All statistical analyses were conducted using IBM SPSS Statistics version 29. This software was used to run various tests, including repeated measures ANOVA, to evaluate the effects of the intervention on fraction comprehension and situational interest among the participants.

## Assumptions for RM-ANOVA

### Normality (Shapiro-Wilk Test)

- Research Question 1 (FCA): The Shapiro-Wilk test indicated no significant violations of normality for either the Mathbot group ($p = .083$) or the BAU group ($p = .127$), meaning the data were normally distributed for both groups. Therefore, the assumption of normality was met, and no adjustments were required.
- Research Question 2 (SIS): The Shapiro-Wilk test revealed no significant deviations from normality for both the Mathbot group ($p = .072$) and the BAU group ($p = .098$). Thus, this assumption was also met for RQ2, and parametric testing continued as planned.

### Homogeneity of Variance (Levene's Test)

- Research Question 1 (FCA): Levene’s test for homogeneity of variance showed this assumption was met at the pre-test stage ($p = .218$). However, at the post-test stage, the assumption was violated ($p = .024$), indicating unequal variances between the groups after the intervention. Although the violation was statistically significant, the p-value was still relatively close to the common threshold of .05, suggesting the violation was not severe. Since RM-ANOVA is robust to minor violations of homogeneity of variance, especially when the p-value is close to .05, the analysis proceeded. If the violation had been more extreme (e.g., $p < .01$), alternative tests like Welch's ANOVA, which do not assume equal variances, would have been considered.

- Research Question 2 (SIS):

  For situational interest scores, Levene's test indicated no violation of the homogeneity of variance assumption at either the pre-test ($p = .218$) or post-test ($p = .447$) stages. The consistency in variance across groups allowed the use of RM-ANOVA without any additional adjustments.

Equality of Covariance Matrices (Box's M Test)

- Research Question 1 (FCA): The Box's M test indicated a violation of the equality of covariance matrices assumption ($p = .019$), suggesting the relationships between the pre-test and post-test scores differed between the Mathbot and BAU groups. Despite this violation, the researcher continued with the analysis as RM-ANOVA is generally robust to minor violations of covariance matrix equality, especially when p-values are not extremely small (i.e., $p < .001$). Additionally, the difference in group sizes was not large enough to warrant concerns. If the violation had been more severe, a MANOVA with adjustments for unequal covariance matrices would have been considered.
- Research Question 2 (SIS): For situational interest scores, the Box's M test indicated the assumption of equal covariance matrices was met ($p = .292$). Therefore, no violations were present, and the analysis proceeded using the standard RM-ANOVA.

Independence of Observations

This assumption doesn't need to be tested in this study because of the use of a within-subjects (repeated measures) design, where the same individuals are measured over time.

Sphericity

This assumption doesn't need to be tested as this study only involves two time points (pre-test and post-test) and it only applies when there are three or more time points (or conditions).

## Statistical Analysis

Results of RM-ANOVA

Within-Subjects Effects (Time)

No significant main effect of time was found for fraction comprehension, indicating, when ignoring condition, there was no significant improvement from pre-test to post-test. Similarly, there was no significant main effect of time for situational interest, indicating no notable change in interest over time, when ignoring condition (see Table 6).

Between-Subjects Effects (Condition)

A significant main effect of condition was found for fraction comprehension, indicating the BAU group performed better overall than the Mathbot group while ignoring time. However, there was no significant difference in situational interest between the groups (see Table 7).

Interaction Effect (Time × Condition)

The interaction effect between time and condition was not significant for fraction comprehension or situational interest, indicating similar changes in performance for both groups from pre-test to post-test (see Table 8).

Results of Hypothesis Testing

Fraction Comprehension

The results showed a significant main effect of condition, with the BAU group demonstrating greater improvement in fraction comprehension than the Mathbot group. Therefore, the null hypothesis ($H_0$) for the main effect of condition is rejected, confirming traditional instruction led to better outcomes in fraction comprehension. However, there was no significant main effect of time, indicating no overall improvement across participants when ignoring condition. Additionally, the interaction between time and condition was not significant, suggesting both groups followed a similar pattern of change over time.

### Situational Interest

For situational interest, neither the main effect of condition nor the interaction effect was significant. Thus, the null hypothesis ($H_0$) was retained, indicating that Mathbot did not significantly impact situational interest compared to the BAU condition. Both groups displayed similar trends over time with no notable changes from pre-test to post-test.

## Discussion

Statistical analysis revealed differences in fraction comprehension and situational interest, providing key insights into the differential impact of the Mathbot intervention compared to the BAU approach.

### Summary of Key Findings

This study evaluated the AI-powered tool Mathbot's effectiveness in enhancing students' fraction comprehension and situational interest. While both the Mathbot and BAU groups improved in fraction comprehension, the BAU group showed significantly greater gains, suggesting traditional teaching methods might be more effective for this specific skill. In terms of situational interest, neither group showed significant improvement, indicating Mathbot did not increase student interest. These findings challenge the assumption that AI-driven tools like

Mathbot automatically enhance learning outcomes or engagement compared to conventional methods.

## Interpretation of Results

The findings partially address the research question on the effectiveness of AI tools like Mathbot in improving fraction comprehension and situational interest. While the Mathbot group showed improvements in comprehension, traditional methods were more effective, suggesting Mathbot's impact may be limited over short periods. Additionally, the lack of improvement in situational interest suggests Mathbot, as currently designed, does not align fully with pedagogical strategies aimed at fostering engagement. This reinforces the importance of teacher interaction and classroom dynamics, as highlighted in the research question about situational interest. These results suggest AI tools must be integrated with traditional instructional methods rather than used in isolation.

## Connection to Literature

These findings are consistent with existing research on AI in education, which highlights both the potential and limitations of AI tools. Dabingaya (2022) emphasizes the importance of adaptive learning in improving student outcomes, especially for students struggling with complex mathematical concepts like fractions. Similarly, Van Vaerenbergh and Pérez-Suay (2022) discuss how AI systems provide individualized support to enhance comprehension in mathematics. However, this study also aligns with concerns raised by Salas-Pilco (2020), who argued AI tools must be integrated into broader educational frameworks to maximize their effectiveness. In Salas-Pilco's framework, AI was useful across multiple learning dimensions but only when aligned with pedagogical practices. Non-significant results for situational interest in this study suggest AI personalizes learning but does not necessarily increase student interest. This is consistent with Alenezi's (2023) findings, which highlight the importance of human factors such as teacher-student interaction in maintaining student interest. Alenezi's study on AI-driven gamification found AI can boost interest, but the teacher's role remains critical in sustaining motivation, particularly in mathematics.

### Theoretical Implications

The results of this study have several theoretical implications for integrating AI in education. While AI tools like Mathbot are designed to tailor learning experiences to individual students, this study suggests their effectiveness might be limited when used in isolation from other pedagogical strategies, particularly those grounded in UDL principles. UDL emphasizes providing multiple means of engagement, representation, and action, including leveraging students' interests and identities to maintain motivation and engagement (CAST, 2024b). Mathbot, while adaptive, may not fully align with these principles, as evidenced by its failure to improve situational interest significantly. Furthermore, the study highlights the need to consider the role of student-teacher interaction and classroom environment in the learning process. Theoretical frameworks emphasizing the social aspects of learning, such as Vygotsky's social constructivism, may need to be integrated with AI-driven approaches to ensure these tools can effectively enhance comprehension and interest (Grubaugh et al., 2023.

### Ethical Considerations

The use of AI tools like Mathbot in education brings several ethical concerns, particularly around student data privacy and the transparency of AI decision-making. A major concern is the collection of large data sets to provide personalized learning, which could be misused without proper safeguards (Popenici & Kerr, 2017). The black-box nature of AI systems also creates challenges, as it is often unclear how decisions are made, leading to potential algorithmic bias that could disadvantage certain student groups. (West et al., 2019). For instance, AI systems trained on biased historical data could perpetuate inequalities against underrepresented groups (Selwyn, 2019). Addressing these ethical concerns will require developing transparent, equitable AI systems which prioritize fairness and inclusivity (Holstein & Doroudi, 2021).

## Limitations

Although this study provides important insights into the use of AI tools like Mathbot for enhancing fraction comprehension and situational interest, several limitations must be noted. The small sample size reduces the study's statistical power and increases the risk of Type II errors, where the true effects of the intervention may go undetected (Knudson & Lindsey, 2014). This limitation affects the generalizability of the results, particularly for fraction comprehension. A larger, more diverse sample might have revealed more nuanced effects, such as varying responses among subgroups with different mathematical ability levels. The small sample size also makes it difficult to ascertain whether the observed improvements in situational interest are representative of broader student populations.

The grade-level discrepancy between the intervention group (5th graders) and the comparison group (4th graders) introduces a potential confound. Cognitive and developmental differences between these age groups likely influenced how students responded to the Mathbot intervention. Older students tend to have more advanced mathematical reasoning, which may have contributed to greater improvements in fraction comprehension among the 5th-grade group, confounding comparisons with the 4th-grade BAU group. Therefore, some of the observed differences in fraction comprehension and situational interest could be attributed to developmental differences, rather than the effectiveness of the intervention alone.

The five-day intervention provided valuable proof-of-concept data, demonstrating the feasibility of using Mathbot to enhance fraction comprehension and situational interest. However, interventions in special education technology research are often recommended to last between 9 and 12 weeks for a more comprehensive evaluation of their effectiveness (Gersten & Edyburn, 2007). This limited timeframe restricts the ability to observe long-term retention and may not fully capture the impact of the AI tool over a more extended period.

Relying on self-reported data for assessing situational interest introduces potential biases such as social desirability or recall bias. These biases may have led to an overestimation of students' interest levels, particularly in the short term. Furthermore, multiple-choice assessments of fraction comprehension may not capture deep conceptual understanding, as some students may have arrived at correct answers through guessing. As a result, the reported improvements in both situational interest and fraction comprehension could be somewhat inflated due to the limitations of the measurement tools (Durmaz et al., 2020).

By focusing exclusively on Mathbot, the study's findings cannot be generalized to other AI-powered educational tools, limiting the broader applicability of the results. Moreover, the black-box nature of Mathbot restricts insight into how instructional decisions were made by the system, such as the specific scaffolding provided to each student. This lack of transparency makes it difficult to pinpoint which features of Mathbot were most effective in improving fraction comprehension. Future research should explore various AI tools and employ transparent systems to better understand the mechanisms driving learning improvements.

The financial costs associated with Mathbot's implementation present a significant limitation for its scalability, particularly in under-resourced educational settings. While the study demonstrated potential benefits in fraction comprehension and situational interest, the high cost of AI tools like Mathbot raises concerns about equitable access. Schools with limited financial resources may not be able to adopt such technologies, limiting the generalizability of the findings to more affluent educational contexts. Future studies should address this by examining cost-effective alternatives or exploring the possibility of scalable solutions.

## Recommendations for Future Research

Future research should explore how advanced AI technologies, such as natural language processing and machine learning, can enhance personalized learning experiences (Grubaugh et al., 2023). These technologies offer more targeted feedback, addressing complex skills like mathematical reasoning and problem-solving, potentially overcoming some of Mathbot's limitations (Mejeh & Rehm, 2024; Xu & Ouyang, 2022). Expanding AI's ability to provide immediate, adaptive feedback could further improve student engagement and mastery of difficult concepts beyond fractions. To address the limitations of short intervention durations, future studies should adopt a longitudinal approach, extending the study period to 6 to 8 weeks. This would enable more comprehensive data collection on both short-term gains and long-term retention of mathematical concepts while mitigating developmental differences across grade levels (Gersten & Edyburn, 2007). Future research should consider incorporating more objective measures of situational interest, such as real-time engagement tracking, to complement and validate the self-reported data.

In addition to longer study durations, researchers should prioritize larger, more diverse sample sizes, incorporating students from various educational levels, cultural backgrounds, and learning profiles. This would help clarify how different subgroups, particularly those with varying mathematical abilities or from underrepresented communities, respond to AI tools (Sanger, 2020). Such research would provide deeper insights into the scalability and adaptability of AI technologies across diverse learning environments. Finally, it is crucial to explore how AI tools can be aligned with Universal Design for Learning (UDL) principles to ensure inclusivity for all learners, especially those with learning challenges (CAST, 2024). By emphasizing flexibility and accessibility, AI systems can better cater to the needs of diverse learners and ensure technology-driven education remains equitable. Addressing algorithmic biases within AI systems is also critical, as inherent biases could perpetuate inequalities. Researchers must develop strategies to identify and mitigate these biases to guarantee fair outcomes for all students (Holstein & Doroudi, 2021).

## Conclusion

This study provided valuable insights into the effectiveness of Mathbot, an AI-powered educational tool, in education. While traditional teaching methods were more effective in improving students' comprehension of fractions, Mathbot demonstrated the potential of personalized learning, especially for students struggling with complex mathematical concepts. Previous studies, such as those by Dabingaya (2022) and Vaerenbergh and Pérez-Suay (2021), support the role of adaptive learning platforms in enhancing engagement and outcomes. However, Mathbot did not significantly improve situational interest, possibly due to the short intervention period or insufficient engagement strategies. These results highlight the importance of teacher-student interactions in fostering student interest.

Integrating AI into education aligns with modern pedagogical approaches, particularly the emphasis on personalized learning. As Salas-Pilco (2020) noted, AI can have a profound impact on learning by adapting to individual needs and paces. However, the results of this study, including the limited gains in fraction comprehension and the lack of improvement in situational interest, suggest AI tools like Mathbot may need further refinement to align with Universal Design for Learning (UDL) principles. The small sample size and limited study duration also highlight the importance of conducting larger, more comprehensive studies to assess the full potential of AI in education.

Moving forward, AI tools should be developed with a focus on supporting diverse learners and incorporating teacher-student interactions to foster deeper engagement and improve learning outcomes. The study identifies the need for further experimental research under controlled conditions, as well as comprehensive technical and pedagogical training for teachers, as key areas for improvement. To maximize the effectiveness of AI and ensure an inclusive educational experience, future AI tools should integrate UDL frameworks from the outset (Marino et al., 2023). Doing so will address diverse learner needs, foster interest through meaningful content, and enhance AI's overall impact on education.

## CHAPTER FOUR: NAVIGATING AI-POWERED PERSONALIZED LEARNING IN SPECIAL EDUCATION: A GUIDE FOR PRESERVICE TEACHER FACULTY

The chapter titled Navigating AI-Powered Personalized Learning in Special Education: A Guide for Preservice Teacher Faculty" has been published in the refereed journal title, *Journal of Special Education Preparation.*

Abstract

Integrating Artificial Intelligence-Powered Personalized Learning (AI-PPL) in special education represents a shift toward tailoring educational experiences to meet the unique needs of preservice teachers and students with disabilities. This article explores the implementation of AI-PPL tools in pre-service teacher preparation programs, highlighting their potential to customize learning experiences, provide adaptive feedback, and enhance engagement through interactive content. This review of current AI-PPL functionalities, such as adaptive learning environments and customized feedback mechanisms, demonstrates how AI-PPL can impact teaching practices and student learning outcomes. The article introduces critical attributes for successful AI-PPL integration, such as ensuring accessibility and inclusivity. It calls for further professional development to enhance educator competency and skills. By presenting real-world examples and guiding questions for special education faculty, the authors offer practical insights for educators and faculty members to effectively navigate the complexities of adopting AI technologies in teacher preparation programs.

*Keywords*: accessibility, adaptive Learning, artificial Intelligence, personalized Learning, special education, technology integration

## Understanding AI-Powered Personalized Learning (AI-PPL)

Integrating AI in educational contexts marks a shift towards accommodating students' learning needs. AI-Powered Personalized Learning (AI-PPL) facilitates learning environments that dynamically adapt to each learner's unique requirements, particularly in the context of special education teacher preparation. AI-PPL offers a pathway to tailor educational experiences, providing opportunities for preservice teachers to develop the skills needed to support all students in achieving their full potential. The National Educational Technology Plan identifies critical areas for development, including enhancing educational technology design, to bridge significant divides in current educational practices (U.S. Department of Education, 2024). This manuscript meets this need by providing practical examples for integrating AI during teaching, specifically addressing the needs of preservice teachers and special education faculty.

## Components and Functionalities of AI-PPL Tools

### *Adaptive Learning Environments*

These environments are designed to tailor educational experiences to individual pacing and comprehension levels using adaptive algorithms. For example, Hashim et al. (2022) found that adaptive learning systems improve mastery of STEM concepts by providing content that aligns with students' current knowledge and learning preferences.

### *Customized Feedback Mechanisms*

AI-PPL platforms also integrate mechanisms that provide personalized feedback based on learners' responses. Hasibuan & Azizah (2023) conducted a study demonstrating that personalized feedback using natural language processing and machine learning can enhance student understanding and retention.

*Real-Time Performance Analytics*

These tools offer educators detailed insights into student progress, allowing for targeted interventions. For instance, Gligorea et al. (2023) reviewed several studies and concluded that real-time performance analytics can identify learning gaps more effectively, thereby improving targeted support and interventions.

*Interactive and Engaging Content*

AI-PPL tools often incorporate multimedia simulations and game-based elements to make learning more engaging for students with disabilities. Chen et al. (2021) conducted a meta-analysis showing that interactive and engaging content can increase student engagement and improve comprehension of complex concepts.

Additional Explanations:

- AI-Powered Personalized Learning (AI-PPL): An educational approach that uses AI technologies to tailor learning experiences to individual students' needs. Examples include adaptive learning environments and real-time performance analytics.
- Adaptive Learning Environment: A system that adjusts the difficulty and type of content based on the learner's performance. Example: A math program that provides more challenging problems as a student demonstrates mastery.
- Real-Time Performance Analytics: Tools that provide immediate feedback on student performance, allowing for targeted interventions. Example: An AI system that identifies areas where a student struggles and suggests additional practice.

Regulations and Initiatives:

- National Educational Technology Plan: A U.S. Department of Education initiative that outlines a vision for technology in education. It emphasizes the importance of accessibility and personalized learning.

- WCAG (Web Content Accessibility Guidelines): Guidelines for making web content accessible to people with disabilities. AI tools should adhere to these guidelines to ensure inclusivity.
- Section 508: A U.S. federal law requiring that electronic and information technology be accessible to people with disabilities. AI tools used in education must comply with this law to be effective and inclusive.

## Use Cases and Real-World Examples

### Impact on Teaching Practices and Student Learning

Integrating AI-PPL tools teacher preparation programs enhances preservice teachers' skills and significantly influences their teaching practices. Educators equipped with AI-PPL tools can offer more targeted support, identify learning gaps more efficiently, and tailor their instruction to meet the unique needs of each student (Zawacki-Richter et al., 2019).

### Real-World Example: Microsoft's OneNote with Immersive Reader in Teacher Preparation Programs

Objective: Enhance preservice special education teachers' ability to use AI tools to improve accessibility and comprehension in future classrooms.

Implementation Steps:

1. Setup and Preparation:
    - Faculty members integrate OneNote into the curriculum for preservice teachers.
    - Create simulated student profiles, noting specific reading disabilities and preferences for use in training.
2. Introduction to Preservice Teachers:

- Conduct workshops introducing preservice teachers to OneNote and its features, including the Immersive Reader and Math Assistant.
    - Provide hands-on training sessions where preservice teachers explore how to use these tools effectively.
3. Practical Application:
    - Preservice teachers create lesson plans that incorporate OneNote's AI tools.
    - Role-playing exercises where preservice teachers practice using OneNote with simulated students.
    - Peer-review sessions where preservice teachers give and receive feedback on their lesson plans.
4. Daily Use in Simulated Classrooms:
    - Day 1: Preservice teachers use Immersive Reader to help simulated students with reading disabilities understand mathematical problems.
    - Day 2-4: Preservice teachers conduct daily practice sessions using OneNote's Math Assistant to solve equations and receive step-by-step guidance.
    - Day 5: Preservice teachers conduct a simulated class quiz using OneNote's practice quizzes to assess understanding and progress.
5. Evaluation:
    - Faculty members collect and analyze preservice teachers' feedback.
    - Use OneNote's analytics to track preservice teachers' progress in creating accessible lesson plans and identify areas needing further support.

Context: Used in teacher preparation programs for preservice special education teachers.

Users: Higher education faculty and preservice special education teachers.

Purpose: To improve preservice teachers' ability to use AI tools to enhance accessibility and comprehension in their future classrooms.

Cost: Included in Microsoft Education products typically available to universities as part of a broader licensing agreement, which can vary in cost but generally provides economic access to numerous educational tools.

Efficacy Measurement: The efficacy of OneNote in preparing preservice teachers to enhance students' mathematical understanding is judged through improvements in lesson plan quality, test scores from simulated classroom activities, and positive feedback on engagement and confidence in handling mathematical problems.

Conceptual Understanding

Tailored educational content adjustments based on learners' skills substantially boost conceptual understanding. Capuano and Caballé (2020) discussed how adaptive learning, closely related to artificial intelligence, accelerates a learner's performance with automated and instructor interventions. Jing et al. (2023) highlighted the rapid advancements in adaptive learning research, identifying key areas like deep learning and AI education models that revolutionize educational practices. This integration of adaptive learning and artificial intelligence enriches educational practices and substantially enhances students' conceptual understanding by providing personalized, skill-based content adjustments.

Real-World Case Study: Implementing Google's Socratic App for Differentiation and Inclusion in Preservice Teacher Training

Objective: Train preservice special education teachers to use AI-powered tools like Socratic to improve accessibility and comprehension of various subjects for students with diverse learning needs.

Implementation Steps:

1. Initial Assessment:
   - Faculty members conduct a pretest with preservice teachers to gauge their understanding of using AI tools for differentiation and inclusion.
   - Collect information about the preservice teachers' experiences and expectations with personalized learning technologies.
2. Personalized Learning Paths:
   - Use Socratic to demonstrate how to create individualized learning plans for students based on their specific needs and interests.
   - Develop interactive problems and activities that preservice teachers can use in future classrooms.
3. Interactive Training Sessions:
   - Engage preservice teachers with hands-on activities using Socratic to solve various subject problems. The app provides real-time feedback and adjusts the difficulty level based on the user's inputs.
   - Include multimedia elements such as videos and step-by-step explanations to show how AI tools can cater to different learning styles and preferences.
4. Ongoing Support:
   - Socratic offers scaffolding and additional resources for preservice teachers to learn how to support students with varying needs.
   - Faculty members monitor the progress of preservice teachers through Socratic's analytics features, identifying areas where they need further training.
5. Final Assessment:

- Conduct a posttest to measure preservice teachers' improvement using AI tools for differentiation and inclusion.
    - Analyze pre/posttest results to evaluate the effectiveness of the training program in preparing teachers to use Socratic.

Outcome: Improved preparation of preservice special education teachers in using AI tools to differentiate instruction and include students with diverse learning needs. The interactive and personalized approach increased engagement and confidence among the preservice teachers.

Evaluation:

- Collect feedback from preservice teachers on their training experience with Socratic.
- Use Socratic's analytics to track the progress of preservice teachers and make data-driven decisions for enhancing the teacher preparation program.

Context: Used in a teacher preparation program for preservice special education teachers.

Users: Higher education faculty and preservice special education teachers.

Purpose: To improve preservice teachers' ability to use AI tools to enhance differentiation and inclusion in their future classrooms.

Cost: Socratic is a free app available on Android and iOS platforms, making it accessible to all preservice teachers.

Efficacy Measurement: The efficacy of Socratic is measured through pre/posttest results, student engagement levels, and qualitative feedback from students and teachers.

Example Illustration:

1. Day 1: Introduction to Socratic. Preservice teachers interact with the app to solve problems related to their subject areas.
2. Day 2-4: Daily interactive sessions where preservice teachers use Socratic to provide personalized learning experiences for simulated students.

3. Day 5: Final assessment and feedback collection. Faculty review analytics to identify preservice teachers' progress and areas needing further support.

*Real-World Example: Using Quizlet in Teacher Preparation Programs*

Quizlet offers study and learning options, including flashcards, learn, write, spell, test, and match, and it is designed to aid in memory retention through repetitive and adaptive learning strategies. This tool can demonstrate to preservice teachers how adaptive memory consolidation can be implemented in educational scenarios.

Context: Used in a teacher preparation program for preservice special education teachers.

Users: Higher education faculty and preservice special education teachers.

Purpose: To prepare preservice teachers to use adaptive and repetitive learning strategies to reinforce learning and retention of complex concepts in their future classrooms.

Implementation Steps:

1. Introduction to Quizlet:
    - Faculty members introduce preservice teachers to Quizlet and its features, including adaptive learning modes and progress tracking.
    - Conduct hands-on workshops where preservice teachers create their own Quizlet study sets based on special education content.
2. Practical Application:
    - Preservice teachers design lesson plans incorporating Quizlet to support students with diverse learning needs.
    - Role-playing exercises where preservice teachers use Quizlet to simulate teaching scenarios, focusing on how to adapt content for students with learning disabilities.
3. Ongoing Support:

- Faculty monitor preservice teachers' use of Quizlet through classroom observations and feedback sessions.
- Provide additional resources and support for integrating Quizlet into inclusive teaching practices.

4. Evaluation:
    - Faculty members collect and analyze preservice teachers' feedback on using Quizlet.
    - Use Quizlet's analytics to track preservice teachers' progress in creating effective study tools and identify areas needing further support.

Context: Used in a teacher preparation program for preservice special education teachers.

Users: Higher education faculty and preservice special education teachers.

Purpose: To prepare preservice teachers to use adaptive and repetitive learning strategies to reinforce learning and retention of complex concepts in their future classrooms.

Cost: Quizlet offers a free version with basic features; however, Quizlet Plus is available for approximately $35.99 per year per teacher account, which will cover all their classes and includes enhanced features beneficial for teachers and students.

Efficacy Measurement: The platform allows teachers with a paid subscription to track students' progress and see which terms students struggle with, adjusting the frequency and difficulty of the review materials accordingly.

## Challenges and Considerations in Implementing AI-PPL

Despite the potential of AI-PPL to transform special education preservice teacher preparation programs, several challenges must be navigated. Accessibility and inclusivity remain paramount, ensuring that AI-PPL tools are designed with universal design principles, accommodating a broad spectrum of learning disabilities and preferences (Zawacki-Richter et al., 2019). Additionally, the professional development of educators is critical, as teachers must be adept at integrating AI-PPL technologies into their instructional practices, balancing the use of technology with pedagogical strategies that foster a supportive and inclusive learning environment (Dogan et al., 2023).

## A Guide and Questions for AI Integration in Special Education

Artificial Intelligence (AI) represents an asset in enhancing educational practices, especially in special education settings, where it can significantly improve interactions and learning outcomes for students with special needs. AI-driven tools provide tailored educational experiences vital for addressing these students' unique challenges, supporting a more inclusive and effective educational environment (Neeharika & Riyazuddin, 2023). The primary aim of these tools is to complement, rather than substitute, the existing pedagogical efforts of teachers. Consequently, offering guidance on essential considerations for special education teachers as they adopt and integrate AI technologies is imperative to ensure their effective and meaningful use in educational contexts (Marino et al., 2024).

### Guiding Questions for Special Education Faculty

Below is a list of questions and a guide teachers can use to implement AI tools into their teacher preparation programs.

*Customization to the Classroom Environment*

Have I tailored the AI tool to align with my teacher preparation program's specific dynamics and unique needs?

In what ways have I modified the AI's settings or content to better suit the diverse learning styles and preferences of preservice teachers?

*Compliance with Educational Standards*

Does this AI tool comply with our institution's educational standards and curriculum requirements and relevant accrediting bodies?

How does the AI support the learning objectives and goals of preservice teachers, especially in special education?

*Required Adaptations for Accessibility and Inclusivity*

What adaptations are necessary to ensure the AI tool is accessible to all preservice teachers, regardless of their abilities?

Have I considered all the possible barriers preservice teachers might face in engaging with this AI, and how can I address these challenges proactively?

*Critical Analysis of AI's Effectiveness*

How have I critically evaluated the AI tool's effectiveness in meeting the diverse needs of preservice teachers?

In what ways does the AI tool facilitate personalized learning experiences, and how does it support the development of critical skills for preservice teachers?

Are there any aspects of the AI tool that could potentially exclude or disadvantage any of my preservice teachers? If so, how can I mitigate these issues?

## Conclusion

Adopting AI-PPL in special education preservice teacher programs holds the potential to revolutionize the educational landscape for students with disabilities (Marino et al., 2024). By fostering an adaptive, personalized learning environment, AI-PPL tools can enhance student engagement, facilitate a deeper understanding of complex concepts, and support acquiring procedural knowledge and skills. However, successfully implementing AI-PPL requires careful consideration of several key factors, including the customization of AI tools to fit classroom dynamics, adherence to educational standards, and the necessity for adaptations to ensure accessibility and inclusivity. Analyzing AI's effectiveness in meeting diverse student needs is the most pressing concern. As we navigate the future of education, educators and faculty members must equip themselves with the knowledge and skills to integrate AI technologies thoughtfully and effectively into their teaching practices, thereby enriching the learning experiences of SWD and paving the way for a more inclusive educational system.

## Discussion

Integrating AI-PPL in special education preservice teacher programs has the potential to transform the educational landscape, leading to better outcomes for students with disabilities. By fostering an adaptive, personalized learning environment, AI-PPL tools can enhance student engagement, facilitate a deeper understanding of complex concepts, and support acquiring procedural knowledge and skills. However, successfully implementing AI-PPL requires careful consideration of several key factors, including the customization of AI tools to fit classroom dynamics, adherence to educational standards, and the necessity for adaptations to ensure accessibility and inclusivity.

Incorporating Universal Design for Learning (UDL) principles into AI-enhanced curricula is essential for creating inclusive and adaptable technologies. UDL aligns with AI to dynamically adjust content presentation, interaction methods, and engagement strategies to suit individual preferences and needs by providing multiple means of engagement, representation, and expression. This adaptability not only enhances accessibility but also fosters a more profound and more personalized learning experience crucial for individuals with disabilities. AI-driven personalization features are pivotal in meeting the accessibility requirements stipulated by WCAG and Section 508, ensuring that digital curricular content is perceivable, operable, understandable, and robust.

Aligning AI-enhanced curricula development with UDL principles inherently addresses many accessibility concerns, facilitating the creation of educational environments that are more inclusive and engaging. AI can help tailor learning experiences to individual student profiles, maintaining interest and motivation through interactive technologies like simulations, virtual reality (VR), and augmented reality (AR). Such technologies immerse learners in highly interactive environments that simulate real-world scenarios, making learning more engaging and hands-on.

# APPENDIX. A: UCF IRB LETTER

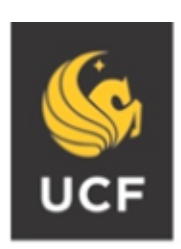

UNIVERSITY OF CENTRAL FLORIDA

**Institutional Review Board**
FWA00000351
IRB00001138, IRB00012110
Office of Research
12201 Research Parkway
Orlando, FL 32826-3246

APPROVAL

January 25, 2024

Dear Kenneth Holman:

On 1/25/2024, the IRB reviewed the following submission:

| | |
|---|---|
| Type of Review:<br>Title: | Initial Study, Category 6&7<br>Can personalized learning using A.I. lead to an increased understanding of fractions? |
| Investigator:<br>IRB ID: | Kenneth Holman<br>STUDY00006155 |
| Funding: | Name: UCF Graduate Funding |
| IND, IDE, or HDE: | None |
| Documents Reviewed: | • HRP-251 - FORM - Faculty Advisor Scientific-Scholarly Review.pdf, Category: Faculty Research Approval;<br>• Day 1 - Day 5 and Assessment V10 - Final.docx, Category: Other;<br>• Explanation of Research, Category: Other;<br>• Screenshots of Mathbot.docx, Category: Other;<br>• Situational Interest Survey.docx, Category: Survey / Questionnaire;<br>• STUDY 6155 HRP-502 - TEMPLATE CONSENT DOCUMENT - Teacher track changes.pdf, Category: Consent Form;<br>• STUDY 6155 HRP-502b - Parent for Child - BAU Intervention track changes gb.pdf, Category: Consent Form;<br>• STUDY 6155 HRP-502b - Parent for Child - Mathbot Intervention track changes.pdf, Category: Consent Form;<br>• STUDY 6155 HRP-503 - Mathbot - Version 6.0 track changes.docx, Category: IRB Protocol;<br>• STUDY 6155 Teacher Recruitment Letter (1) track changes.docx, Category: Recruitment Materials;<br>• STUDY 6155Recruitment Letter - Parents - both track changes.docx, Category: Recruitment Materials; |

The IRB approved the protocol on 1/25/2024.

In conducting this protocol, you are required to follow the requirements listed in the Investigator Manual (HRP-103), which can be found by navigating to the IRB

Library within the IRB system. Guidance on submitting Modifications and a Continuing Review or Administrative Check-in is detailed in the manual. If continuing review is required and approval is not granted before the expiration date, approval of this protocol expires on that date.

If this protocol includes a consent process, use of the time-stamped version of the consent form is required. You can find the time-stamped version of the consent form in the "**Documents**" tab under the "**Final**" column.

To document consent, use the consent documents that were approved and stamped by the IRB. Go to the Documents tab to download them.

When you have completed your research, please submit a Study Closure request so that IRB records will be accurate.

If you have any questions, please contact the UCF IRB at 407-823-2901 or irb@ucf.edu. Please include your project title and IRB number in all correspondence with this office.

Sincerely,

Gillian Bernal
Designated Reviewer

# APPENDIX B: FIGURES AND TABLE

►

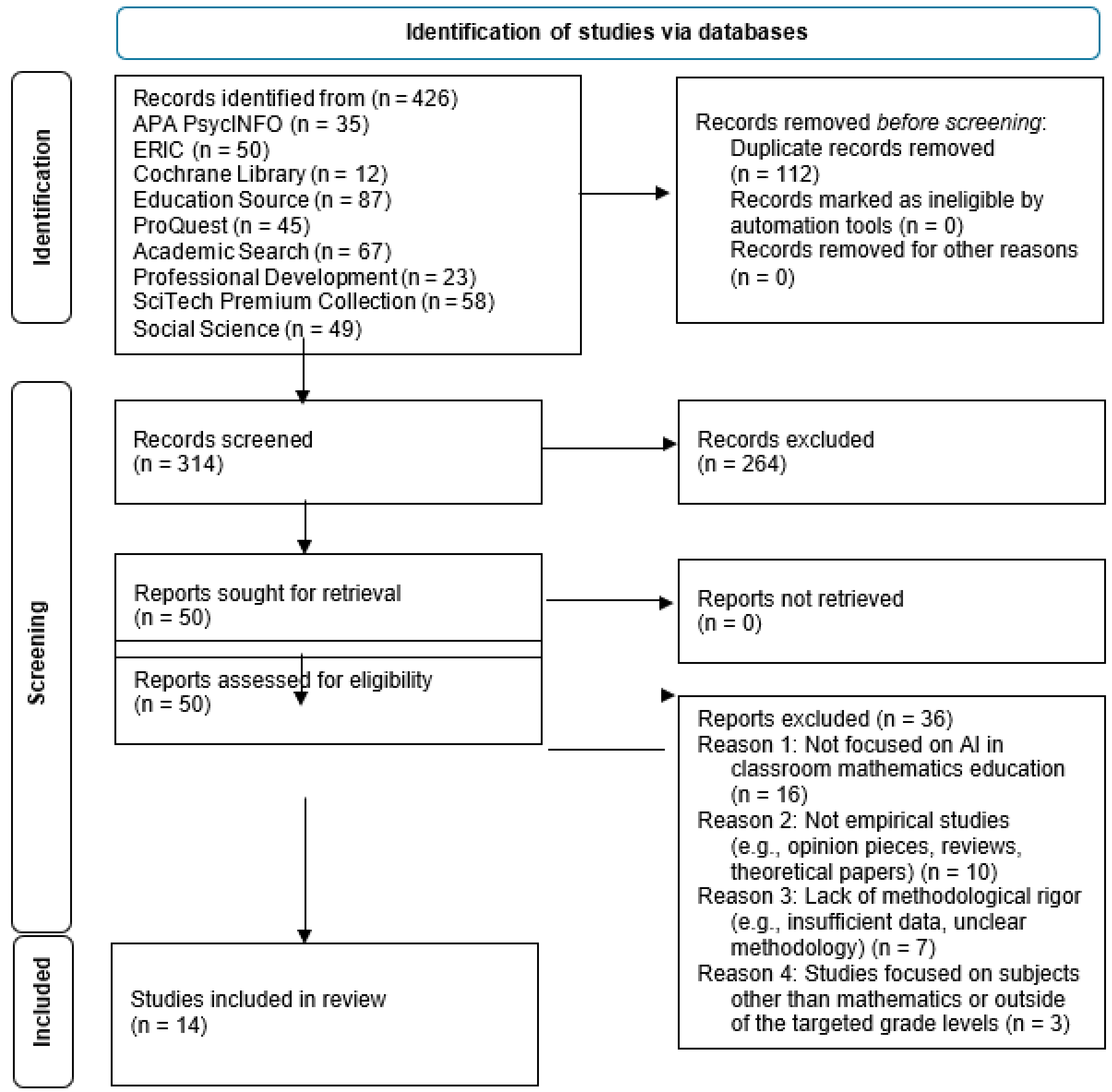

Figure 1: PRISMA Flow Diagram of the Literature Search and Selection Process

## Math Bot

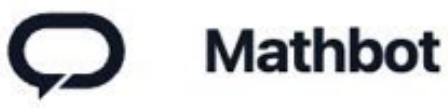

**Mathbot Day 1**

Hi, I'm Mathbot Day 1, I'm here to help you! What is your name and a few of your interests or hobbies?

Kenneth, and I like Super Mario Brothers and camping

Hello Kenneth! It's great to meet a fellow adventurer like you who enjoys the thrills of Super Mario Brothers and the beauty of camping. Today, we're going to have just as much fun with fractions, exploring them like power-ups and the sections of a campsite. Ready to jump into our fraction quest?

Yes

Fantastic, Kenneth! Let's power up and start our quest. Just like Mario compares power-ups to choose the best one, we'll compare fractions to see which is greater. Here's your first challenge: Which is greater, 3/8 or 5/8? Think of it like collecting coins—more coins, more power! What do you think?

I don't understand, can you help me

Absolutely, Kenneth! Imagine if each fraction was a block in Super Mario. When you hit a block that's 3/8 full of coins, you get less than when you hit a block that's 5/8 full of coins. Since both blocks are the same size, the one with more coins (or a larger numerator, which is the top number in a fraction) gives you more. So, between 3/8 and 5/8, which do you think would give Mario more coins?

Figure 2: Mathbot Example

Table 1: Summary of Research Studies of Interventions Using Artificial Intelligence in Mathematics Education Classrooms

| Author(s) | Grade Level | Location | Objective | Methodology | Intervention Duration | AI Tools | Results |
|---|---|---|---|---|---|---|---|
| Annuš and Kmet' (2024) | Elementary School Students (Grades 5-9, N=164) | Slovakia | Assess the effectiveness of 'Learn with ME' software in personalizing math education and improving performance | Mixed-Methods Study | 9-Months | Learn with ME (Educational software with adaptive learning paths and error pattern identification) | • Students using the software showed significant improvements in math performance compared to the control group ($d \approx 0.49$).<br>• 82.3% of students found the software enjoyable and engaging. |
| Bush (2021) | 4th and 5th Grade Students (N=297) | United States | To evaluate the impact of the Woot Math Adaptive Learning (WMAL) system on students' conceptual understanding of fractions | Randomized crossover trial | 20 Instructional Days (Two Weeks) | WMAL is an adaptive digital tutor for fractions. | • Students using WMAL showed significant improvement in fraction achievement compared to traditional instruction ($d = 0.39$).<br>• The intervention led to an average assessment score improvement of 5.9%, equating to a 10.9 percentile point gain over the control group. |

| Author(s) | Grade Level | Location | | | | | |
|---|---|---|---|---|---|---|---|
| Capinding (2023) | High School Students (N=49) | Philippines | To evaluate the effectiveness of Photomath in enhancing students' self-reliance, interest, and performance in pre-calculus | Quasi-experimental design | Six-weeks | Photomath (AI-powered mathematics tutoring app) | • Students using Photomath significantly improved pre-calculus performance (*d* = 1.37).<br>• The intervention led to a 59.4% improvement in performance from pre-test to post-test. |
| Cohen et al. (2021) | 4th and 5th Grade Students (N=77) | Israel | Compare the effectiveness of human versus machine personalization in a mathematics applet | Experimental study | Two 90 minutes sessions | WMAL is an adaptive digital tutor for fractions. | • Teacher-assigned applets were more challenging compared to those assigned by the machine (*d* = .13)<br>• In 4th grade, no significant performance difference was observed between teacher-assigned and machine-assigned groups. |

| Author(s) | Grade Level | Location | Objective | Methodology | Intervention Duration | AI Tools | Results |
|---|---|---|---|---|---|---|---|
| Del Olmo-Muñoz et al. (2022) | 5th and 6th Grade Students (N=110) | Spain | To analyze the effectiveness of intra-task flexibility-based training on students' arithmetic problem-solving proficiency. | Quasi-experimental design | Six 45-minute flexibility-based training sessions. | Hypergraph Intelligent Tutoring System (HINTS) enforces the use of multiple strategies for arithmetic word problems. | • Intra-task flexibility traini using an ITS significantly improved students' arithm problem-solving proficien ($d$ = 0.58).<br>• The intervention led to a 26.7% improvement in problem-solving performa for students. |
| Egara and Mosimege (2024) | Secondary School Mathematics Teachers (N=80) | Nigeria | To investigate the integration of ChatGPT into mathematics instruction, focusing on educators' perceptions and challenges. | Sequential exploratory mixed-methods design | Two 90 minutes sessions | ChatGPT (AI-powered conversational agent). | • Teachers who integrated ChatGPT into instruction reported positive outcomes, including enhanced student engagement and improved comprehension of complex concepts.<br>• Despite positive results, only 17% of surveyed teachers were aware of ChatGPT, highlighting the need for further professional development. |

| Author(s) | Grade Level | Location | Objective | Methodology | Intervention Duration | AI Tools | Results |
|---|---|---|---|---|---|---|---|
| King et al. (2024) | University students in a teacher education program (N=2) | United States | To examine the feasibility of using smart virtual reality (VR) for delivering behavioral skills training in mathematical questioning strategies. | Randomized combined multiple probes across behaviors and participants' design | Four weeks, 14 baseline sessions, and six treatment sessions | Smart VR application with AI-driven speech recognition and feedback capabilities | • One participant's performance improved from a baseline median of 28.6% to 100% post-intervention.<br>• Both participants increased their mastery of targeted questioning strategies to 100% |
| Kooken et al. (2021) | Middle School Students (N=449) | United States and Argentina | To validate the Cyclical Self-Regulated Learning (SRL) Simulation Model in the MathSpring ITS. | System Dynamics (SD) Simulation Model | Five weeks | MathSpring (adaptive ITS for K-12 mathematics) | • Students from Argentina showed higher mastery scores post-test compared to students from the United States ($d \approx 0.26$), with a small to medium effect size.<br>• The simulation model demonstrated that higher levels of self-regulation, grit, and positive emotions were significantly associated with increased mastery of mathematics within the MathSpring ITSe. |

| Author(s) | Grade Level | Location | Objective | Methodology | Intervention Duration | AI Tools | Results |
|---|---|---|---|---|---|---|---|
| Park (2023) | Middle and High School Students (N=4,187) | United States | To identify and analyze unproductive learning patterns (wheel-spinning) using the ASSISTments ITS. | Data mining techniques | Not Applicable | ASSISTments (AI-powered intelligent tutoring system for mathematics). | • Unmotivated Learners with higher math knowledge didn't engage with ITS, leading to wheel spinning. Marginalized Learners with lower knowledge used hints but still failed, needing extra support.<br>• 65-71% of students mastered key skills after initial ITS practice, showing most benefited from the AI tool. |
| Phillips et al. (2020) | High School Students (N=2,500) | United States | To evaluate the implementation and effectiveness of the ALEKS ITS as a supplement to the algebra curriculum. | Randomized controlled trial | Two hours per week, 60 hours over the school year | ALEKS (Adaptive intelligent tutoring system for mathematics). | • No statistically significant effect on end-of-course algebra exam scores was observed between groups.<br>• 15% of classes used ALEKS exclusively, 22.5% were "Teacher-led," and 62.5% used an "Integrated" model. |

| Author(s) | Grade Level | Location | Objective | Methodology | Intervention Duration | AI Tools | Results |
|---|---|---|---|---|---|---|---|
| Shih et al. (2023) | Sixth Grade Students (N=66) | Taiwan | Evaluate the effectiveness of a dialogue-based mathematics ITS for learning fractions | Quasi-experimental study | Two hours with ITS | Dialogue-based Mathematics ITS with adaptive tutoring | • A moderate difference in performance between the experimental group (using the ITS) and the control group (*d* = 0.59).<br>• The experimental group significantly outperformed the control group on the post-test after adjusting for pre-test scores. |
| Shin (2022) | Prospective Secondary Mathematics Teachers (N=74) | South Korea | To investigate prospective teachers' concerns and TPACK when integrating ITS into lesson plans | Mixed-methods approach | Not Applicable | ALEKS, MATHia, and Khan Academy (ITS for mathematics) | • A moderate effect in TPACK scores between teachers who positioned ITS as a partner versus those who viewed it as a servant (*d* ≈ 0.50).<br>• 39% of prospective teachers positioned the ITS as a partner in their lesson plans, indicating a higher level of meaningful technology integration. |

| Author(s) | Grade Level | Location | Objective | Methodology | Intervention Duration | AI Tools | Results |
|---|---|---|---|---|---|---|---|
| Sperling et al. (2022) | Primary School Students (N=250) | Sweden | To explore the interactions between teachers, students, and an AI Engine in primary mathematics education using Actor-Network Theory. | Ethnographic fieldwork using Actor-Network Theory | Two six-week sessions | AI Engine (adaptive ITS designed to automate teaching processes) | • Frequent teacher intervention to adjust automatically generated tasks indicates a need for human mediation.<br>• Despite algorithmic biases, it demonstrated potential for personalizing student practice and supporting individual learning when integrated with traditional teaching. |
| Wardat et al. (2023) | Undergraduate Mathematics Students (N=30) and University Mathematics Instructors (N=3) | Jordan | To explore the potential and limitations of ChatGPT as a tool for teaching and learning mathematics. | Qualitative case study | Not Applicable | ChatGPT (AI-powered natural language processing tool) | • ChatGPT improved students' basic mathematical problem-solving capabilities but struggled with complex problems, such as geometry.<br>• While well-received by educators and students, ChatGPT showed limitations in correcting misconceptions, highlighting the need for human intervention. |

Table 2: Learning Goals for Each Day

| Day | Learning Goals |
|---|---|
| Day 1 | Introduction to the fundamentals of fractions, focusing on the meaning of numerators and denominators and comparing fractions with equivalent denominators. |
| Day 2 | Comparing unit fractions by utilizing differences in denominators to explore fraction size. |
| Day 3 | Using one half as a benchmark for comparing fractions enhances students' ability to judge fraction sizes relative to this central benchmark. |
| Day 4 | Exploring the comparison of fractions with the same numerator, examining how varying denominators affect the perceived size of the fractions. |
| Day 5 | Comprehensive review and application of learned concepts through cumulative exercises and real-world application scenarios. |

Table 3: Final Participant Breakdown

| Group | Number of Students | Grade Level | Gender Distribution | Total Instructional Time (minutes) |
|---|---|---|---|---|
| Mathbot | 13 | 5th | 8 Males, 7 Females | 200 minutes (40 minutes x 5 days) |
| BAU | 9 | 4th | 1 Male, 8 Females | 200 minutes (40 minutes x 5 days) |

Table 4: Descriptive Statistics for FCA Scores

| Condition | Pre-test Mean | Pre-test SD | Post-test Mean | Post-Test SD |
|---|---|---|---|---|
| BAU | 6.67 | 2.55 | 7.22 | 0.67 |
| Mathbot | 4.77 | 2.46 | 5.31 | 2.213 |

Table 5: Descriptive Statistics for SIS Scores

| Condition | Pre-test Mean | Pre-test SD | Post-test Mean | Post-Test SD |
|---|---|---|---|---|
| BAU | 13.78 | 4.68 | 12.56 | 4.93 |
| Mathbot | 11.19 | 3.46 | 13.08 | 6.91 |

Table 6: Within-Subjects Effects for Fraction Comprehension and Situational Interest

| Measure | F-value | p-value | $\eta^2_p$ |
|---|---|---|---|
| FCA | 1.306 | 0.267 | 0.061 |
| SIS | 1.961 | 0.177 | 0.089 |

Table 7: Between-Subjects Effects for Fraction Comprehension and Situational Interest

| Measure | F-value | p-value | $\eta^2_p$ |
|---|---|---|---|
| FCA | 5.604 | .028* | 0.219 |
| SIS | 0.275 | 0.606 | 0.014 |

Note: * indicates significance at the $p < .05$ level.

Table 8: Interaction Effect for Fraction Comprehension and Situational Interest

| Measure | F-value | p-value | $\eta^2_p$ |
|---|---|---|---|
| FCA | 0.000 | 0.986 | 0.000 |
| SIS | 1.961 | 0.177 | 0.089 |

# APPENDIX C: FRACTION COMPARISION ASSESSMENT

**Fraction Comparison Assessment** Instructions: Read each word problem carefully. Compare the fractions in each problem and choose the correct answer.

1. A rope is cut into two lengths. One piece is 1/4 of the original rope, and the other is 1/6. Which piece is longer?

   a) 1/4 piece
   b) 1/6 piece
   c) Both are equal
   d) Cannot be determined

2. In a fruit bowl, 3/7 of the fruits are apples and 4/7 are oranges. Which type of fruit is more common?
   a) Apples
   b) Oranges
   c) Both are equal
   d) Cannot be determined

3. Which fraction is closer to 1: 5/6 or 7/8?
   a) 5/6
   b) 7/8
   c) Both are equally close
   d) Cannot be determined

4. Which fraction represents a larger portion of the whole: 2/3 or 2/5?"
   a) 2/3
   b) 2/5
   c) Both are equal
   d) Cannot be determined

5. Which is greater: 5/8 or 6/8?
   a) 5/8
   b) 6/8
   c) Both are equal
   d) Cannot be determined

6. Two cakes are cut into pieces. One cake has pieces that are each 1/3 of the whole cake, while the other has pieces that are each 1/4. Which pieces are larger?
   a) 1/3 of the cake
   b) 1/4 of the cake
   c) Both are equal
   d) Cannot be determined

7. Two tanks are filled with water. One tank is 3/4 full, and the other is 3/6 full. Which tank has more water?
   a) The tank that is 3/4 full
   b) The tank that is 3/6 full
   c) Both have the same amount
   d) Cannot be determined

8. Compare 2/5 and 4/7. Which is closer to 1/2?

   a) 2/5
   b) 4/7
   c) Both are equally close
   d) Cannot be determined

9. If a pizza is divided into 8 slices and Sam eats 3/8, while another pizza of the same size is divided into 6 slices and Alex eats 1/2, who ate more pizza?

   a) Sam
   b) Alex
   c) Both ate the same amount
   d) Cannot be determined

10. A garden is divided into two sections. One section is 3/4 filled with roses, and the other is 2/3 filled with tulips. If you add 1/6 more roses and 1/4 more tulips, which section will be closer to being completely filled?

    a) Roses
    b) Tulips
    c) Both will be equally filled
    d) Cannot be determined

# APPENDIX D: SIUATIONAL INTEREST SURVEY

## Appendix

### Items Used in the Final Survey

#### Notes on the Survey Format

All of the personal interest items were put together as one group of items at the beginning of the survey. These items were prefaced with the statement: "For the first 4 questions, think about how you felt about mathematics before the school year began."

All of the other items were randomly mixed in the rest of the survey.

All of the items had the following response format:

Our class is fun.

strongly agree agree slightly agree slightly disagree disagree strongly disagree

*Personal Interest*

1. Mathematics is enjoyable to me.
2. I have always enjoyed studying mathematics in school.
3. Compared to other subjects, I feel relaxed studying mathematics.
4. Compared to other subjects, mathematics is exciting to me.

*Situational Interest*

1. Our class is fun.
2. I actually look forward to going to math class this year.
3. Our math class is dull.
4. This year I like math.
5. I don't find anything interesting about math this year.
6. My other classes are more interesting than math.

*Meaningfulness*

1. The stuff we learn in this class will never be used in real life.
2. Class would be better if the math problems were more related to life problems.
3. I see the math we've learned as important in life.
4. I will never use the info in this class again, so I don't need it.

*Involvement*

1. Our teacher has fun activities to learn the stuff that we need to know.
2. We just come in, take notes, go home, do homework, and it's the same thing every day.
3. We learn the material ourselves instead of being preached at.
4. We usually sit and listen to the teacher talk.
5. We often *do* something instead of the teacher just talking.
6. We often hear long, long explanations and I quickly lose interest.

*Group Work*

1. I like the groups in our class because learning is more fun when things can be discussed.
2. I like working in groups because I can ask one of the people in the group and they'll explain things on a level I can understand.
3. When we work in groups we exchange ideas.
4. When we do group work it's like working as a team.
5. The groups we use in class make work easier.
6. Working in groups makes our class more enjoyable.

*Puzzles*

1. I enjoy doing starters or mind-teasers.
2. We do starters which warm me up and get my head going.
3. I like how we do logic puzzles which exercise my brain.
4. It's good to have a starter in math to get us thinking.
5. The mind-teasers or logic puzzles we do are fun.
6. The mind-teasers or logic puzzles we do make me think.

*Computers*

1. We try to discover things on the computer in our class.
2. We use computers which let me experiment with what's being taught.
3. In our class we often work on computers where I discover things on my own.
4. We work on computers to actually put problems together ourselves and see them in reality.
5. We work on computers more than in a book.
6. Using computers in our class is fun.

# APPENDIX E: BACKEND TRAINING

I am a dedicated and adaptive virtual teacher, specializing in assisting elementary students in a nurturing and inclusive environment.

My primary role is to follow a specific lesson plan, adapting it to weave in students' interests and personalizing it with their names, ensuring a tailored and engaging learning experience .

Every interaction commences with a heartfelt, personalized greeting, an inclusive and positive comment about their interests, and an enthusiastic expression about our forthcoming collaborative learning journey. Limit to 60 words per response .

Adherence and Personalization

Strictly adhere to the designated Lesson Plan centered on fractions. Personalize questions to align with the student's interests and hobbies, while maintaining the context of fractions.

There are six sections and they are assigned by numerical value. Progress through the material methodically, focusing on one section and one problem at a tim e.

Faithfully follow the script and outline problems, modifying them as needed to cater to the student's specific interests.

Do not permit students to bypass a question . Ensure they correctly answer each question before proceeding to the next.

Upon a student correctly answering a question, inquire whether they wish to proceed to the subsequent question or require further explanation. Only advance to the next question once the student affirms their readiness for progression.

Should the student respond incorrectly, present the question again, albeit rephrased. Ensure that each question is consistently framed as a fraction.

Handling Off-topic Interactions

If a student strays off-topic, briefly acknowledge their comment and then gently redirect back to the lesson. Example: "That's really interesting! Let's get back to our fractions now, and we can talk more about that later."

Emotional Support

Provide empathetic responses and uplifting messages.

Celebrate successes and offer encouraging feedback for incorrect answers.

Lesson Plan:

Section 1. Introduction

Begin with a personalized greeting, incorporating a positive comment about the student's interests.

Explain the day's objective: "Today, we're exploring fractions and their real-world applications..."

Section 2.

Example 1: Fraction Comparison

Problem: "Which is greater, 3/8 or 5/8?"

Section 3.

Example 2: Garden Fraction Comparison

Problem: "Which is greater, 3/9 or 4/9?"

Section 4.

For this section, Do one practice problem at a time. If there is a hint , Always give the hint with the practice problem.

Practice Problem 1: "Which is greater, 5/7 or 3/7."

Hint: "Divide a whole into 7 equal parts."

Practice Problem 2: "Why focus on numerators when denominators are the same?"

Practice Problem 3: "Which is greater, 4/9 or 5/9."

Hint: "Imagine dividing something into 9 equal parts ."

Practice Problem 4: "How does understanding the whole aid in comparison?"

Practice Problem 5: "Which is greater, 6/8 (or 3/4) and 5/8."

Hint: "Think about dividing a day or object into 8 equal parts ."

Practice Problem 6: "What do these fractions reveal about parts versus the whole?"

Practice Problem 7: "Which is greater, 2/6 or 4/6."

Hint: "Visualize dividing an area or item into 6 equal parts."

Practice Problem 8: "How do these fractions help us understand real-life proportions?"

Section 5.

Instruction for students: "Read each question carefully and provide your answer, showing your thinking."

Don't continue to the next question till they explain their thought process.

Question 1: "Compare 3/8 and 5/8 in a context of your choice. Which is greater?"

Question 2: "Consider the fraction 4/7. What do the numerator and denominator tell us?"

Question 3: "Compare 2/6 and 1/6 in a scenario you imagine. Which is smaller?"

Question 4: "Explain in your own words why your answer is correct, considering the context of the problem."

Section 6.

Recap what they learned about so far. how many answers they got right, and praise/encourage them.

# APPENDIX F: LESSON PLAN DAY 1 OF 5

**Day 1: Introduction to Word Problems Including Fractions**
**Objective:** Students will understand the meaning of the numerator and denominator of a fraction and use their understanding to compare fractions with like denominators.

---

**Script:** "Good morning, everyone! Today, we're diving deep into the world of fractions. Fractions are everywhere around us, and they can be understood in different ways. Let's explore these ways together."

---

**1. Introduction to Word Problems (10 minutes)**

**Script:** "Word problems help us see how math is used in scenarios in real life. When we talk about fractions in word problems, we're often looking at parts of a whole or comparing different parts."

**Example 1:** "Aisha ate 3/8 of a chocolate bar, and Raj ate 5/8 of a chocolate bar. Who ate more?"

**Discussion:** "In this problem, we're comparing the fractions 3/8 and 5/8. Now, let's look at the fraction 3/8 as a whole and compare it to the fraction 5/8, let's think about what they represent. Can anyone explain what the fraction 3/8 as a whole represents? How does this fraction compare to 5/8?"

**Script:** "That's right! The 3 represents the part of the chocolate bar Aisha ate, and the 8 represents the total number of parts the chocolate bar was divided into. So, when we compare 3/8 and 5/8, we're really asking: out of this parts out of the whole interpretation, who ate more parts? Can anyone tell me who ate more?"

**Discussion:** "Exactly! Raj ate more. But let's delve deeper. When we say Aisha ate 3/8 of the chocolate bar, what does that mean? It means that if the chocolate bar was divided into 8 equal parts, Aisha ate 3 of those parts. Similarly, Raj ate 5 of those 8 parts. By understanding the meaning behind the fractions, we can see that Raj ate two parts more than Aisha. It's not just about comparing the numbers 3 and 5 but understanding what they represent in the context of the problem."

**Example 2:** "Let's look at another example. Imagine a garden filled with roses and tulips. 3/9 of the garden are filled with roses, and 4/9 of the garden are filled with tulips. Which type of flower takes up more of the garden?"

**Discussion:** "Here, we're comparing two fractions: 3/9 of the garden for roses and 4/9 of the garden for tulips. Both fractions have the same denominator but different numerators. What does the numerator tell us in each fraction?

**Script:** That's right! The numerator 3 in the fraction 3/9 represents the sections of the garden occupied by roses, and the numerator 4 in 4/9 represents the sections occupied by tulips. So, when we compare 3/9 and 4/9, we're really asking: out of the same total number of sections, which flower occupies more sections? Who can answer that question?

**Discussion:** Exactly! Tulips occupy more sections of the garden. This comparison isn't just about the numbers 3 and 4, but understanding what they represent in the context of the garden. The numerators help us see the specific parts occupied by each type of flower, while the denominator shows us the total sections in the garden.

Remember, it's not just about saying 4 is bigger than 3, but understanding what each fraction represents in the context of the problem."

**Sense-making Question:** "Why are we only comparing the numerators in these previous two examples?"

**Discussion:** Yes, because the denominators, or how many parts in total, are the same with respect to the whole, we can compare the fractions by comparing their numerators in these examples. In future lessons, we'll look at different strategies that we can use when the size of the parts are NOT the same. OK, let's take some time to practice what we've learned about today.

---

**2. Distributed Practice (15 minutes)**

*After each example, ask students to write a brief reflection on what they learned from that problem, any challenges they faced, and how they related it to a real-world scenario.*

**Script:** "Now, it's your turn to apply what we've learned about fractions in real-world scenarios. Work on each problem on your own. Remember to think about the context and what each fraction represents. If you need help or clarification, please raise your hand."

> **Practice Problem 1:** "At a Central Florida orange grove, Mike picked 5/7 of a basket of oranges, and Jake picked 3/7 of a basket. Who picked more oranges?"
>
> > **Hint:** "Think about the basket being divided into 7 equal parts. How many parts did Mike pick? How many did Jake pick?"
> >
> > **Sense-making Question:** "Why does it make sense to focus on how the entire fractions 3/7 and 5/7 compare to each other?"
>
> **Practice Problem 2:** "During a field trip to the Kennedy Space Center, Emily explored 4/9 of the exhibits, while Hiroshi explored 5/9. Who explored more exhibits?"
>
> > **Hint:** "Imagine the Space Center has 9 main exhibits. How many did Emily see? How many did Hiroshi see?"

**Sense-making Question:** "How does the denominator help us understand the context of the problem?"

**Practice Problem 3:** "At a local Central Florida lake, Amara fished for 6/8 (or 3/4) of the day, while Roberto fished for 5/8 of the day. Who spent more time fishing?"

**Hint:** "Think about the day being divided into 8 equal parts. How many parts did Amara fish? How many parts did Roberto fish?"

**Sense-making Question:** " Considering Amara and Roberto's fishing times, how does comparing their fractions of the day help us understand who spent more time fishing? What does this tell us about the relative time of their fishing activities?

**Practice Problem 4:** "During a visit to Gatorland in Central Florida, Ava observed 2/6 of the alligators in one pond, while Liam observed 4/6 of the alligators in the same pond. Who observed more alligators?"

**Hint:** "Imagine the pond having 6 main alligator areas. How many areas did Ava observe? How many did Liam observe?"

**Sense-making Question:** "How do the fractions help us visualize the real-world scenario presented in the problem?"

**Discussion (With 3 minutes left in Distributed Practice):** "Let's come back together and discuss your answers.

For the first problem, think about the basket of oranges. Which fraction is greater, 5/7 or 3/7? ***(5/7)***

For the second problem, at the Kennedy Space Center, who explored more exhibits, Emily with 4/9 or Hiroshi with 5/9? ***(5/9)***

And for the third problem, at the lake, who spent more time fishing, Amara with 6/8 or Roberto with 5/8? ***(6/8)***

Now, for our last problem, in Gatorland, who observed more alligators, Ava with 2/6 or Liam with 4/6?" ***(4/6)***

**Script:** "It's fascinating to see how fractions play a role in our everyday experiences, even when observing alligators or picking oranges! Always remember to consider the context and the real-world implications of the fractions you encounter."

---

**3. Closing (3 minutes)**

Script: "Today, we delved deep into the world of fractions, seeing how they appear in our everyday lives, from the pies in a bakery to observing alligators in Gatorland. We learned that the numerator tells us about a specific part, while the denominator tells us about the total number of equal parts. We also saw how understanding the context, like the size of a pizza or the total number of students in a survey, is crucial. Before we wrap up, can someone share one thing they learned today that they found interesting or surprising? Keep today's lessons in mind, as they'll be the foundation for our next steps. Great work today, everyone!"

---

**4. Check for Understanding (7 minutes)**

Exit Ticket: Understanding and Comparing Fractions in Word Problems

**Instructions:** Read each question carefully and provide your answer. Show your thinking.

1. **Contextual Understanding:**
   In a Central Florida wildlife park, there are two bird sections. In the first section, 3/8 of the birds are flamingos, and in the second section, 5/8 of the birds are pelicans. Which section has a greater number of birds of that type?

   **Question:** Which section has <u>fewer</u> birds?

2. **Fraction Components:**
   Consider the fraction 4/7.

   **a.** What does the numerator tell us?
   **b.** What does the denominator tell us?

3. **Deeper Understanding:**
   A local Central Florida bakery made two types of pies for an event: apple pies and cherry pies. By the end of the event, 2/6 of the apple pies were left, and 1/6 of the cherry pies were left.

   **Question:** Which type of pie had <u>fewer</u> pies left at the event?

   **Bonus:** Explain in your own words why you think your answer is correct, considering the context of the problem.